\newif{\ifarxiv}
\newif{\ifremarks}

\arxivtrue  
\remarkstrue  

\newcommand\arxor[2]{\ifarxiv{#1}\else{#2}\fi}

\ifremarks
\newcommand{\remarktb}[1]{{\renewcommand{\bfdefault}{b}{\color[RGB]{0,150,0}{\textbf{#1}}}}}
\newcommand{\remarkcb}[1]{{\renewcommand{\bfdefault}{b}{\color[RGB]{0,0,150}{\textbf{#1}}}}}
\newcommand{\JAM}[1]{{\renewcommand{\bfdefault}{b}\color{olive}[\textbf{JAM: #1}]}}
\fi
\providecommand{\remarktb}[1]{\ignorespaces}
\providecommand{\remarkcb}[1]{\ignorespaces}
\providecommand{\JAM}[1]{\ignorespaces}

\documentclass[
 reprint,
 amsmath,
 amssymb,
 aps,
 prl,
 nofootinbib,
 longbibliography,
 superscriptaddress,
 nobalancelastpage,
 preprintnumbers
]{revtex4-2}


\usepackage[a4paper,vmargin=2cm,hmargin=1.5cm]{geometry}

\usepackage[utf8]{inputenc}
\usepackage{comment}
\usepackage[pdftex]{graphicx}
\usepackage{wasysym}
\usepackage{amsfonts}
\usepackage{ifsym}

\usepackage[dvipsnames]{xcolor}
\usepackage{color}
\usepackage{graphicx}
\usepackage{dcolumn}
\usepackage{bm}
\usepackage{hyperref}

\hypersetup{plainpages=false}
\hypersetup{pdfpagemode=UseOutlines}
\hypersetup{bookmarksnumbered=true}
\hypersetup{bookmarksopen=true}
\hypersetup{pdfstartview=FitH}
\hypersetup{colorlinks=true}
\hypersetup{citecolor=[rgb]{0 .4 0}}
\hypersetup{urlcolor=[rgb]{.4 0 0}}
\hypersetup{linkcolor=[rgb]{0 0 .5}}

\newcommand{\namedref}[2]{\hyperref[#2]{#1~\ref*{#2}}}

\newcommand{\appref}[1]{\namedref{Appendix}{#1}}

\newcommand{\figref}[1]{\namedref{Figure}{#1}}

\makeatletter
\def\mr@ignsp#1 {\ifx\:#1\@empty\else #1\expandafter\mr@ignsp\fi}%
\newcommand{\multiref}[1]{\begingroup
\xdef\mr@no@sparg{\expandafter\mr@ignsp#1 \: }%
\def\mr@comma{}%
\@for\mr@refs:=\mr@no@sparg\do{\mr@comma\def\mr@comma{,\,}\ref{\mr@refs}}%
\endgroup}
\makeatother
\renewcommand{\eqref}[1]{(\multiref{#1})}

\allowdisplaybreaks
\def\etal.{et\penalty50\ al.}
\usepackage{xspace}

\newcommand*{\ie}{i.\,e.\@\xspace}
\makeatletter\newcommand*{\etc}{%
    \@ifnextchar{.}%
        {etc}%
        {etc.\@\xspace}%
}\makeatother

\def\clap#1{\hbox to 0pt{\hss#1\hss}}

\usepackage[mathbold,autobold,greekcaps,greeklower]{mathfixs}

\usepackage{empheq}
\newlength{\widefboxpadding}
\newcommand*\widefbox[1]{\fbox{\hspace{\widefboxpadding}#1\hspace{\widefboxpadding}}}

\newcommand{\software}[1]{\texttt{#1}\xspace}
\newcommand{\mathematica}{\software{Mathematica}}

\RequirePackage[extdef]{delimset}
\makeatletter
\providecommand{\brkleft}[1][r]{\begingroup\def\dlm@use{\delim(.}%
\if r#1 \def\dlm@use{\delim(.}\fi%
\if s#1 \def\dlm@use{\delim[.}\fi%
\if c#1 \def\dlm@use{\delim\{.}\fi%
\if a#1 \def\dlm@use{\delim<.}\fi%
\expandafter\endgroup\dlm@use}
\providecommand{\brkright}[1][r]{\begingroup\def\dlm@use{\delim.)}%
\if r#1 \def\dlm@use{\delim.)}\fi%
\if s#1 \def\dlm@use{\delim.]}\fi%
\if c#1 \def\dlm@use{\delim.\}}\fi%
\if a#1 \def\dlm@use{\delim.>}\fi%
\expandafter\endgroup\dlm@use}
\makeatother

\DeclareMathOperator{\tr}{Tr}

\newcommand{\gym}{g_{\scriptscriptstyle\mathrm{YM}}}

\newcommand{\dd}[2][]{\mathinner{d\ifx#1\empty\else{^#1}\fi#2}}

\begin{document}

\preprint{DESY-24-085}

\title{Wilson Loops with Lagrangians:\texorpdfstring{\\}{ }Large Spin OPE and Cusp Anomalous Dimension Dictionary}

\author{Till Bargheer}
\email{till.bargheer@desy.de}
\affiliation{Deutsches Elektronen-Synchrotron DESY, Notkestr.~85, 22607 Hamburg, Germany}
\author{Carlos Bercini}
\email{carlos.bercini@desy.de}
\affiliation{Deutsches Elektronen-Synchrotron DESY, Notkestr.~85, 22607 Hamburg, Germany}
\author{Bruno Fernandes}
\email{up201706002@edu.fc.up.pt}
\affiliation{Centro de Fisica do Porto e Departamento de
Fisica e Astronomia, Faculdade de Ciencias da Universidade do Porto, Porto 4169-007, Portugal}
\author{Vasco Gon\c{c}alves}
\email{vasco.dfg@gmail.com}
\affiliation{Centro de Fisica do Porto e Departamento de
Fisica e Astronomia, Faculdade de Ciencias da Universidade do Porto, Porto 4169-007, Portugal}
\author{Jeremy Mann}
\email{jeremy.mann@kcl.ac.uk}
\affiliation{Department of Mathematics, King’s College London, Strand, London, WC2R 2LS, UK}

\begin{abstract}
In the context of planar conformal gauge theory, we study five-point correlation functions between the interaction
Lagrangian and four of the lightest single-trace, gauge-invariant scalar primaries. After
performing two light-cone OPEs, we express this correlator in terms of
the three-point functions between two leading-twist spinning operators
and the Lagrangian. For finite values of spin, we compute these
structure constants in perturbation theory up to two loops in
$\mathcal{N}=4$ super Yang--Mills theory.
Large values of spin are captured by null polygon kinematics, where we
use dualities with null polygon Wilson loops as well as factorization
properties to bootstrap the universal
behavior of the structure constants at all loops. We find explicit
maps that relate the Lagrangian structure constants with the
leading-twist anomalous dimension. From
the large-spin map, we recover the cusp anomalous dimension at strong
and weak coupling, including genus-one terms.
\end{abstract}


\maketitle

\section{Introduction}

The operator product expansion (OPE) encodes the data of a conformal
field theory (CFT) in its four-point correlation functions. Capturing
all CFT data requires infinitely many four-point functions. Iterating
the OPE, this infinity of data can in turn be packaged in
higher-point functions of the simplest operators. This is the
philosophy of the multi-point
bootstrap~\cite{Bercini:2020msp,Bercini:2021jti,Antunes:2021kmm,Buric:2022ucg,Kaviraj:2022wbw,Poland:2023vpn,Antunes:2023kyz,Poland:2023bny,Harris:2024nmr},
which trades an infinity of data for a larger functional complexity.

In null polygon limits, this complexity reduces, and the conformal
bootstrap is enhanced by dualities with Wilson loops, both at
four~\cite{Alday:2013cwa} and higher
points~\cite{Bercini:2020msp,Bercini:2021jti}. While null squares and
pentagons allow for no finite conformal cross ratios, null hexagons
are complicated functions of three variables. Here, we consider a
sweet spot: The null square limit of a five-point function, which has
a single finite cross ratio.

We will focus on the correlation function of four single-trace
lightest scalar operators and the interaction Lagrangian in planar
conformal gauge theories. Such correlators yield integrands for scalar
operators~\cite{Eden:2011we}. In the null square limit, they probe the
quantum corrections to null-square Wilson
loops~\cite{Alday:2011ga,Alday:2012hy,Alday:2013ip}, and in particular
were used to compute the full four-loop cusp anomalous dimension for
$\mathcal{N}=4$ super Yang-Mills (SYM) and QCD~\cite{Henn:2019swt}.

By studying the Lagrangian correlation function via the conformal
bootstrap, we translate all its properties to its OPE constituents: The
three-point functions of two leading-twist spinning operators and the
Lagrangian. At finite values of spin, we compute these structure
constants at weak coupling and connect them, via conformal
perturbation theory, to leading-twist anomalous dimensions. For large
values of spin, we find an exact inversion formula that leads to
direct maps between the null-square correlator to the structure
constant~\eqref{eqFDerivation}. Using
conformal perturbation theory at large spin, we obtain an even simpler
map between these structure constants and the cusp anomalous
dimension~\eqref{eqCusptoB}.

\begin{figure}
\centering
\includegraphics{FigOverview3}
\clap{\hspace{-8.9cm}\raisebox{0.05cm}{\eqref{eqFDerivation}}}%
\clap{\hspace{-8.9cm}\raisebox{1.06cm}{\eqref{eqInversion}}}
\caption{Correlation function of four scalar operators
(black dots) and one Lagrangian operator
(crossed circles) in the null square limit, and the maps
that we obtain.}
\label{fig:overview}
\end{figure}
%

\section{Perturbative Data}
\label{secSingle}

We consider five-point functions of one primary scalar operator
$\mathcal{O}(x)$ and four of the lightest scalar operators~$\phi$ of the theory. For
example, in $\mathcal{N}=4$ SYM these would be the $20^\prime$
operators $\phi_j \propto \tr(y_j \cdot \Phi(x_j))^2$. It is convenient to extract a
space-time dependent prefactor of the five-point correlator
\begin{multline}
\langle \phi_1\dots\phi_4\mathcal{O}(x_5)\rangle
\equiv
\left(
\frac{1}{x_{12}^{2}x_{34}^{2}}
\right)^{\Delta_\phi}
\left(
\frac{x_{14}^2}{x_{15}^2x_{45}^2}
\right)^{\Delta_{\mathcal{O}}/{2}}
\mspace{-20mu} \times \\ \times
\prod_{i=1}^{n_\mathcal{O}} (y_i\cdot y_{i+1})\times G_\mathcal{O}(u_i)
+(\text{other})
\,,\label{eq:prefactor}
\end{multline}
where $n_\mathcal{O}=4,5$ depending on whether the fifth operator carries
R-charge or not, and (other) refers to other
$y$-contractions that will be subleading in all the limits we
consider. In this way, $G_{\mathcal{O}}(u_i)$
becomes a function of five cross ratios
\begin{equation}
    u_i = \frac{x_{i,i+1}^2x_{i+2,i-1}^2}{x_{i,i+2}^2x_{i+1,i-1}^2}\,, \quad i=1,\dots,5\,,
\end{equation}
where we identify the points $(x_1,\dots,x_5)$ periodically. Two
particular correlators will be important for us: The correlation
function of five light operators ($G_\phi$), and the five-point function of four light correlators and one Lagrangian ($G_\mathcal{L}$).

To study these correlators, we will consider two light-like
OPEs~\cite{Ferrara:1974nf} between the lightest operators, as depicted
on the right of \figref{fig:overview}. The
leading behavior under this Lorentzian OPE is controlled by the
exchange of leading-twist (twist-two) operators in the OPE
decomposition:
\begin{equation}
    G_\mathcal{O}(u_i) =\sum_{J_1,J_2,\ell}\mathcal{F}(u_i)\times C(J_1)C(J_2)C_{\mathcal{O}}(J_1,J_2,\ell)\,,
    \label{eqStartRelation}
\end{equation}
where $C(J)$ are the structure constants of one leading-twist operator
with spin $J$ and two lightest scalars operators, while
$C_\mathcal{O}(J_1,J_2,\ell)$ are the three-point functions of two
leading-twist spinning operators and the operator $\mathcal{O}(x)$. The
quantum number $\ell = 0,1,2,\dots,\text{min}(J_1,J_2)$ labels the
tensor structures of three-point functions with two spinning
operators~\cite{Costa:2011dw}. Meanwhile, $\mathcal{F}$ is the
theory-independent conformal block worked out
in~\cite{Bercini:2020msp} and recalled in\arxor{~\eqref{eq:LCBlock}}{ Appendix~A}.

In principle, using the integrability formalism for spinning
operators~\cite{Basso:2015zoa,Bercini:2022gvs}, it is possible to
compute the structure constants $C_\phi$ at any order in perturbation
theory. However, the structure constants $C_\mathcal{L}$ are not on
the same integrability footing: Despite some tree-level
results~\cite{Eden:2023gso}, it is presently not clear how to
systematically consider superdescendants like the Lagrangian in the
integrability formalism.

In perturbation theory, we can explicitly evaluate the
correlator $G_\mathcal{L}$ \cite{Eden:2011we,Bercini:2024pya}, and
use it to extract the perturbative data $C_\mathcal{L}$.
We extracted thousands of OPE coefficients up to two loops, contained
in the attached \mathematica file. This data could be useful to
develop future integrability formulations. We were able to identify a
pattern and write the tree-level data as
\begin{multline}
    C_\mathcal{L}^{(0)}(J_1,J_2,\ell)
    =
    {2\frac{2\,J_1!}{\sqrt{(2J_1)!}}\frac{2\,J_2!}{\sqrt{(2J_2)!}}}
    \times\\\times
    \brk[s]*{
    (-1)^\ell \binom{J_1+J_2}{\ell-1}
    +\sum\limits_{m=0}^{\ell-1}\binom{J_1}{m}\binom{J_2}{m}
    }\,.
\end{multline}

Since the Lagrangian is exactly marginal,
conformal perturbation theory relates the two-point function of two
operators with the three-point functions of the two
operators and the Lagrangian in a differential equation~\cite{Zamolodchikov:1987ti}. For the case of
spinning operators this was worked out in~\cite{Sen:2017gfr} to be
\begin{equation}
    \frac{\partial \gamma(J)}{\partial \lambda} = \sum_{\ell=0}^{J}\frac{C_\mathcal{L}(J,J,\ell)}{1+J-\ell}\,,
    \label{CPTFinal}
\end{equation}
where $\Delta_J = 2+J+\gamma(J)$ is the dimension of the leading-twist
operator. A remarkable feature of this anomalous dimension is that in
any planar gauge theory, it develops logarithmic scaling at large
values of spin~\cite{Korchemsky:1988si,Gubser:2002tv}:
\begin{equation}
    \gamma(J) \simeq f(\lambda)\ln(J) + g(\lambda)\label{eqGammaToCusp}
    \,,
\end{equation}
where $f(\lambda)$ and $g(\lambda)$ are the cusp and
collinear anomalous dimensions respectively, with
$\lambda=\gym^2N/(4\pi)^2$, where $\gym$ is the Yang-Mills coupling.
Below we
evaluate~\eqref{CPTFinal} at large values of spin, obtaining a map
between the large-spin Lagrangian structure constants and the
ubiquitous cusp anomalous dimension.

\section{Null Square}

We approach
the null square limit of the five-point function with the Lagrangian
by first taking
$x_{12}^2,x_{34}^2\to 0$ (or $u_1,u_3\to0$), projecting into
leading-twist operators. Next, we take $x_{23}^2\to0$ (or $u_2\to0$),
which we find projects both to large spin $J_i$ and
large polarization $\ell$. Finally, we take $x_{14}^2\to0$,
which makes the two values of spin approach each other, $J_2\to J_1$.

The intuition is that once we
create a null square inside a five-point
function, the OPE decomposition starts developing four-point-like
features. Four-point functions have only one spinning operator flowing
in the OPE channel, and this is exactly what the leading term of the
five-point function reproduces. We make this precise in
\arxor{\appref{appSquareBlock}}{Appendix~A}, via the so-called ``Casimir trick''
introduced
in~\cite{Alday:2015eya,Alday:2015ewa,Simmons-Duffin:2016wlq}, and
systematized for higher-point functions in~\cite{Kaviraj:2022wbw}. In
the end, the five-point block in the null square limit becomes a
simple Bessel-Clifford function
\begin{multline}
\mathcal{F}(u_i) = \left(u_1 u_3\right)^{1+\gamma(J)}\left(u_2 u_4u_5\right)^{\frac{\Delta_\mathcal{L}}{2}} 2^{2J-2+\gamma(J)+\frac{\Delta_\mathcal{L}}{2}}\times\\
\times \pi^{-1/2} J^{\frac{1+\Delta_\mathcal{L}}{2}} \mathcal{K}_{\Delta_\mathcal{L}/2}\left(Ju_2(J+j_1 u_4 +j_2 u_5)\right)\,,
 \label{eqNullBlockMain}
\end{multline}
where $\mathcal{K}_n(z)=z^{-n/2}K_{n}(2\sqrt{z})$, and we introduced the variables
\begin{equation}
J^2 = \frac{J_1^2 + J_2^2}{2}\,, \quad j_1 = J_1-\ell\,, \quad j_2 = J_2-\ell\,.
\label{eqCasimirVar}
\end{equation}
The null square limit is described by all these variables being large
($J,j_1,j_2 \to \infty$) with $J \gg j_1,j_2$, while the ratio $r =
{j_2}/{j_1}$ is finite. This single finite quantum number is
associated with the single cross-ratio $x$ that remains finite in the
null square limit:
\begin{equation}
    x = \frac{u_4}{u_5} = \frac{x_{13}^2x_{25}^2x_{45}^2}{x_{15}^2x_{24}^2x_{35}^2}\,.
\end{equation}

From here onward we will consider Born-level normalized quantities,
which we denote by $\hat{G} = G/G^{(0)}$, in order to make our
statements universal and independent of the prefactor choices such
as~\eqref{eq:prefactor}.

Conformal symmetry implies that null square correlators must factorize into two terms~\cite{Alday:2011ga},
\begin{equation}
    \hat{G}_\mathcal{L}(u_1,u_2,u_3,u_4,u_5) = \hat{G}_{4}(u,v) \times \hat{F}(x)
    \,,
    \label{eqFact}
\end{equation}
which are invariant under cyclic permutations of the null square,
$(x_1,x_2,x_3,x_4) \to (x_2,x_3,x_4,x_1)$ with $x_5$ fixed. This imposes
\begin{align}
    \hat{G}_4(u,v) = \hat{G}_4(v,u) \quad \text{and} \quad \hat{F}(x) = \hat{F}\left(\frac{1}{x}\right)\,,
    \label{eqCyclicityGF}
\end{align}
where $u = u_1 u_3$ and $v = u_2$ are four-point cross ratios.

The first term $\hat{G}_4(u,v)$ is the null four-point function of the lightest
operators, which captures all the divergences of the Lagrangian
correlator, and depends on the four-point cross ratios $u$ and $v$.
The second term $\hat{F}(x)$ is a finite
function of the remaining finite cross ratio.

Thus our bootstrap problem is: Can we fix the universal behavior of the structure
constants such that the Lagrangian correlator factorizes into the square symmetric
functions~\eqref{eqFact}? To start answering this question, we use the
explicit expression for the conformal blocks~\eqref{eqNullBlockMain}
to write the null square correlator as
\begin{multline}
    \hat{G}_\mathcal{L} =\left(u_2^3u_4u_5\right)\int dJ\,dj_1\,dj_2\,
    (u_1 u_3)^{\frac{\gamma}{2}}2^{2+\gamma}J^3 \hat{C}(J)^2
    \,\times \\ \times
    \hat{C}_{\mathcal{L}}(J_1,J_2,\ell)
    \,\mathcal{K}_{2}\brk{Ju_2(J+j_1 u_4 +j_2 u_5)}
    \,,
    \label{eqBootstrapStart}
\end{multline}
where we factored out the tree-level large-spin
scaling of the structure constants:
\begin{subequations}
\label{eqCHat}
\begin{align}
C(J_1) = C(J_2) &\simeq 2^{-J}\pi^{1/4}J^{1/4}\times\hat{C}(J)\,, \\
C_\mathcal{L}(J_1,J_2,\ell)&\simeq 8 \times \hat{C}_\mathcal{L}(J_1,J_2,\ell)\,.
\end{align}
\end{subequations}
The tree-level behavior~\eqref{eqCHat} shows the physics
of these structure constants: $\hat{C}(J)$ is large and captures
the divergent part $\hat{G}_4$ of the correlator in the null square limit. On
the other hand, the structure constant
$\hat{C}_\mathcal{L}(J_1,J_2,\ell)$ is finite and controls the finite part of the correlator $\hat{F}(x)$. We expect that
it only depends on the finite ratio~$r$,
\begin{equation}
    \hat{C}_\mathcal{L}(J_1,J_2,\ell) = \hat{C}_\mathcal{L}\left(\frac{J_2-\ell}{J_1-\ell}\right) \equiv \hat{C}_\mathcal{L}(r)
    \,.
    \label{eqAssumeChat}
\end{equation}
Indeed, we can prove this to be true, using
a five-point null square inversion formula\arxor{ (see \appref{appInversion})}{, see Appendix~B}.

Assuming the simple dependence~\eqref{eqAssumeChat} allows us to
integrate~\eqref{eqBootstrapStart} in one of the two variables $j_i$,
resulting in the following factorized expression for
the null square correlator:
\begin{multline}
    \hat{G}_\mathcal{L}(u_i) = \underbrace{
    	\int_{0}^\infty dJ\,  2^{2+\gamma}J\hat{C}(J)^2u^{\gamma/2} v K_{0}(2J\sqrt{v})
    }_{\hat{G}_4(u,v)} \times \\
    \times \underbrace{
    	\int_{0}^{\infty} dr\,  \frac{x}{(r+x)^2}\hat{C}_\mathcal{L}(r)
    }_{\hat{F}(x)}\,.
    \label{eqGFactorized}
\end{multline}
The first term is \textit{exactly} the same as the null square
four-point function of lightest operators considered
in~\cite{Alday:2013cwa} and therefore automatically obeys the
cyclicity~\eqref{eqCyclicityGF}. The
invariance under $x \to 1/x$ of the function $\hat{F}(x)$ is also
automatically satisfied, provided that $\hat{C}_{\mathcal{L}}(r) =
\hat{C}_{\mathcal{L}}(1/r)$. Physical structure constants must have
this property, since inverting the ratio $r$ is the same as swapping
the spins $J_1 \leftrightarrow J_2$. Thus, the map between
$\hat{F}(x)$ and the Lagrangian structure constants is simply
\begin{empheq}[box=\widefbox]{equation}
\hat{F}(x) = x\int_{0}^{\infty}dr\,\frac{\hat{C}_{\mathcal{L}}(r)}{(x+r)^2}\,.
\label{eqFDerivation}
\end{empheq}

We can invert this map by noticing that the
right hand side is the derivative of the Cauchy kernel, whose
inversion is well understood in terms of its discontinuities.
Therefore, one can write the structure constants in terms of
discontinuities of $F(x)$:
\begin{empheq}[box=\widefbox]{equation}
    \eval*{r\frac{d}{dr}\hat{C}_{\mathcal{L}}(r)}_{r\geq{0}}
    =
    \frac{\text{Disc}}{2\pi i}\hat{F}(-r)\,,
 \label{eqInversion}
\end{empheq}
where we used the fact that physical structure constants
$\hat{C}_\mathcal{L}(r)$ must be regular at physical values of spins and polarization ($r\geq0$).

\section{Weak and Strong Coupling}

Both weak and strong coupling results for the
function $\hat{F}(x)$ have been computed in $\mathcal{N}=4$ SYM. We can use these results
together with our map~\eqref{eqInversion} to compute the structure
constants $\hat{C}_\mathcal{L}$ in these regimes. At weak
coupling, the first orders of $\hat{F}(x)$ were computed
in~\cite{Alday:2011ga,Alday:2012hy,Alday:2013ip,Henn:2019swt},
\begin{align}
    \hat{F}^{(0)}(x) &= 1\,, \nonumber\\
    \hat{F}^{(1)}(x) &= -6\zeta_2-2H_{00}\,,\nonumber\\
    \hat{F}^{(2)}(x) &=
    24\zeta_2H_{-1-1}-12\zeta_2H_{-10}+24\zeta_2H_{00}
    \nonumber\\ &
    +8H_{-1-100}-4H_{-1000}+12H_{0000}-4H_{-200}
    \nonumber\\ &
    -12\zeta_2H_{-2}+8\zeta_3H_{-1}-4\zeta_3H_0+107\zeta_4\,,
\label{eq:hatFexpansion}
\end{align}
where $H_{a}\equiv H_{a}(x)$ are harmonic
polylogarithms~\cite{Remiddi:1999ew}, recalled in
\arxor{\appref{app3Loops}}{Appendix~C},
where we also collect the three-loop and genus-one contributions of
$\hat{F}(x)$.\footnote{The
higher-genus contributions to the cusp anomalous dimension
start at four loops, but due to its derivative relation with this
quantity, $\hat{F}(x)$ features genus-one terms already at
three loops.}

The discontinuities of the harmonic polylogarithms appearing in the
perturbative expansion of $\hat{F}(x)$ can be easily evaluated using
the \software{HPL} package~\cite{Maitre:2005uu} for \mathematica,
resulting in the following expression for the weak coupling structure
constants:
\begin{align}
    \hat{C}_{\mathcal{L}}^{(0)}(r) &= 1\,,\nonumber\\
    \hat{C}_{\mathcal{L}}^{(1)}(r) &= -4\zeta_2-2H_{00}\,,\nonumber\\
    \hat{C}_{\mathcal{L}}^{(2)}(r) &=
    56\zeta_4-4\zeta_3H_{0}+8\zeta_2H_{2}+12\zeta_2H_{00}
    \nonumber\\ &
    +8H_{210}+4H_{200}+4H_{30}+12H_{0000}\,,
    \label{eqPertubativeB2}
\end{align}
where $H_a\equiv H_a(r)$, and
the three-loop and genus-one corrections are written in
\arxor{\appref{app3Loops}}{Appendix~C}.
In practice, the discontinuity fixes all but the constant term. This in
turn can be determined by performing the explicit integration
in~\eqref{eqFDerivation}, and matching with the $\hat{F}(x)$
expansion~\eqref{eq:hatFexpansion}.\footnote{The integrals of harmonic polylogarithms can also be
trivially done using the package \software{HPL}~\cite{Maitre:2005uu}.}

Even though the individual harmonic polylogarithms have a branch point
at $r=1$, the particular combination appearing in the weak-coupling
expansion of $\hat{C}_\mathcal{L}(r)$ is real and single-valued for
physical values of spins and polarizations ($r>0$). This is not true
for the unphysical region $r<0$, where $\hat{C}_\mathcal{L}(r)$ has a
logarithmic branch cut.

At strong coupling, the leading behavior of the function $\hat{F}(x)$
is known~\cite{Alday:2011ga},
\begin{equation}
    \hat{F}(x) = \frac{x}{(x-1)^2}\left(\frac{(x+1)}{(x-1)}\frac{\log{x}}{2}-1\right)\sqrt{\lambda} + \dots\,.
\end{equation}
Using the inversion formula~\eqref{eqInversion}, we can compute
the leading term of the structure
constant at strong coupling,
\begin{equation}
    \hat{C}_{\mathcal{L}}(r) = \frac{r}{2(1+r)^2}\sqrt{\lambda} + \dots
    \,.
    \label{eqBstrong}
\end{equation}

\section{Wilson Loops and Amplitudes}

In $\mathcal{N}=4$ SYM, $n$-point correlation functions of $20^\prime$
operators in the limit where their insertions approach the cusp of a
null polygon are dual to both null polygonal Wilson loops and MHV
gluon scattering amplitudes~\cite{Alday:2010zy,Eden:2010zz}. In
particular, in the five-point null pentagon limit:
\begin{align}
    \lim_{x_{i,i+1}^2\to0}\hat{G}_\phi
    = (\widehat{\text{MHV}}_5)^2 \,,
\end{align}
By promoting this relation to supercorrelation functions and
superamplitudes, one obtains that the correlation function of four
$20^\prime$ correlators and one Lagrangian, when the points approach
the cusps of a null pentagon is dual to (the top component of) the
NMHV scattering amplitude~\cite{Eden:2011yp,Eden:2011ku}
\begin{align}
    \lim_{x_{i,i+1}^2\to0}\hat{G}_\mathcal{L}
    = \widehat{\text{MHV}}_5 \times \widehat{\text{NMHV}}_5 \,,
\end{align}
For five points, the NMHV amplitude is the parity
conjugate of the MHV amplitude,\footnote{This implies that one is the complex conjugate of
the other. Parity-odd terms (imaginary) are important to establish the
duality at integrand level, however they stem from a total derivative
and integrate to zero.} thus in
the null pentagon limit both correlators are identical
\begin{equation}
    \lim_{x_{i,i+1}^2\to0} \hat{G}_\phi
    = \lim_{x_{i,i+1}^2\to0} \hat{G}_\mathcal{L} = \langle \widehat{W}_5 \rangle\,,
    \label{eqDuality}
\end{equation}
which immediately implies $\hat{C}_\mathcal{L}=\hat{C}_\phi$, that
is\footnote{As the null square, the null pentagon limit
is also governed by the regime of large spin and large polarization.
The difference is that for the pentagon there are neither finite
cross-ratios, nor finite ratios among the quantum numbers $J_i,\ell$, see
\arxor{\appref{appNullPent}}{Appendix~D}.}
\begin{multline}
    \hat{C}_{\mathcal{L}}(J_1,J_2,\ell) =
    \\
    \mathcal{N}(\lambda)\,
    e^{-\frac{f(\lambda)}{4}(\log{\ell}^2+2\log{2}\log{(J_1J_2)})-\frac{g(\lambda)}{2}\log{\ell}}
    \,.
    \label{eqCNullPent}
\end{multline}

The story is completely different when we consider the null square
limit of these five-point correlators. As pointed out
in~\cite{Alday:2011ga}, the duality with Wilson loops continues to hold even if
one adds an extra operator at finite distance to the null square
configuration
\begin{equation}
    \lim_{x_{1,2}^2,x_{2,3}^2,x_{3,4}^2,x_{1,4}^2\to0} \hat{G}_{\mathcal{L}} = \langle \widehat{W_4\mathcal{L}}\rangle\,.
\end{equation}

One can recast this duality as an equation for
$\hat{F}(x)$ by using that Lagrangian correlators are
obtained from a derivative with respect to the coupling,
\begin{equation}
    \frac{\partial}{\partial \lambda} \log{\langle \hat{W}_4\rangle}= 8\int dx_5\,  \frac{x_{13}^2x_{24}^2}{x_{15}^2x_{25}^2x_{35}^2x_{45}^2}\hat{F}(x)\,\,.
    \label{eqLogWtoF}
\end{equation}
where the space-time prefactor arises from the Born-level ratio
$\langle\phi_1\dots\phi_4\mathcal{L}(x_5)\rangle^{(0)}/\langle\phi_1\dots\phi_4\rangle^{(0)}$.

\section{Cusp Anomalous Dimension}

The UV cusp divergences of the Wilson loop are controlled by the cusp
anomalous dimension. In principle, one can match the divergences
appearing on both sides of the relation~\eqref{eqLogWtoF} to compute
this quantity. In practice, this is done with the help of the
functional $\mathcal{I}$ formulated in~\cite{Alday:2013ip} and
recalled below,\footnote{The factor 8 arises from
the fact that all our quantities are Born-level normalized except for the
cusp anomalous dimension. We can drop this factor by considering
$\hat{f}(\lambda)$, but we refrain from doing that to avoid confusion.}
\begin{equation}
    \frac{\partial f(\lambda)}{\partial \lambda}=\mathcal{I}[8\hat{F}(x)]\,,
    \label{eqCuspToF}
\end{equation}
where one is instructed to first expand the function $\hat{F}(x)$
around small
values of $x$,\footnote{The cusp singularities will arise when $x_5$
approaches the cusp points $x_i$, which correspond to $x\to 0$
or $x\to\infty$. Due to the symmetry $x\to 1/x$ of this function, both
regimes map to the small $x$ asymptotic.}
and then act with the linear functional on individual terms as
\begin{equation}
    \mathcal{I}[x^{p}] = \frac{\sin{\pi p}}{\pi p}\,.
\end{equation}

Starting from the conformal perturbation theory
relation~\eqref{CPTFinal}, we propose an alternative and more
explicit map. It relates the three point function
$\hat{C}_\mathcal{L}$ with the cusp anomalous dimension simply as
\begin{empheq}[box=\widefbox]{equation}
\frac{\partial f(\lambda)}{\partial \lambda}=8 \hat{C}_{\mathcal{L}}(1)\,.
\label{eqCusptoB}
\end{empheq}

The large-spin limit of the sum~\eqref{CPTFinal} is dominated by the
region where spins and polarizations are of the same order. Therefore,
we can trade the sum over
polarizations by an integral and replace the structure constants
by their large-spin and polarization behavior~\eqref{eqAssumeChat}.
Since the sum runs over structure constants of identical spins, the ratio $r$ becomes
one, and $\hat{C}_\mathcal{L}(1)$ becomes a constant that can be
factored out of the integral. The integral is then trivial and
evaluates to $\log{J}$. Matching the
log-divergent terms on both sides of
equation~\eqref{eqGammaToCusp} yields the
map~\eqref{eqCusptoB}.

We verify this result by recovering
the known values of the cusp anomalous dimension at strong and weak
coupling, including genus-one terms:
At strong coupling, replacing $r=1$ in~\eqref{eqBstrong} and using the
map~\eqref{eqCusptoB} yields the leading term of the cusp anomalous
dimension: $f(\lambda) \simeq 8\sqrt{\lambda}$.
Similarly at weak coupling\arxor{, by evaluating~\eqref{eqPertubativeB2}
and~\eqref{eqPertubativeB3} at $r=1$}{}, we recover the four-loop anomalous
dimension~\cite{Henn:2019swt}:
\begin{align}
8\hat{C}_{\mathcal{L}}(1) &= 8 - 32\zeta_2\lambda+528\zeta_4\lambda^2-\Big(64\zeta_3^2+1752\zeta_6+\nonumber\\
&+\frac{1}{N^2}(1152 \zeta_3^2+2976 \zeta_6)\Big)\lambda^3\,.
\end{align}

The map between three-point functions and the cusp anomalous
dimension~\eqref{eqCusptoB} is simpler than the map~\eqref{eqCuspToF}
previously considered
in the literature. However since the structure
constants and the function $F(x)$ are also related to each other
via~\eqref{eqFDerivation} we must have the following consistency
condition for the structure constant:
\begin{align}
\mathcal{I}\left[ x\int_{0}^{\infty} dr \, \frac{\hat{C}_{\mathcal{L}}(r)}{(x+r)^2}\right] = \hat{C}_{\mathcal{L}}(1) \label{eqTrivial}\,.
\end{align}
Unfortunately, this is \emph{not} a bootstrap equation for
$\hat{C}_{\mathcal{L}}(r)$. One simple way to see this, is to expand
this function as a power series and note that the
relation~\eqref{eqTrivial} acts trivially on each polynomial term
\begin{align}
\mathcal{I}\left[ x\int_{0}^{\infty} dr \, \frac{r^{p}}{(x+r)^2}\right]  =\frac{\pi p}{\sin{\pi p}} \mathcal{I}\left[x^p \right] = 1\,,
\label{eq:Ionrp}
\end{align}
and therefore~\eqref{eqTrivial} is trivially satisfied for any
function $\hat{C}_{\mathcal{L}}(r)$. One might be worried that the
expression above is only valid for $|p|<1$, and that
$\hat{C}_\mathcal{L}(r)$ has no regular expansion around $r=0$.
However, using the physical
properties of the structure constants, \ie invariance under swapping the
spins $\hat{C}_{\mathcal{L}}(r) = \hat{C}_{\mathcal{L}}(1/r)$ and
regularity around $r=1$ (where we recover the cusp anomalous dimension) we can analytically continue
this result for any~$p$, see \arxor{\appref{appTrivial}}{Appendix~E}.

\section{Conclusion}

Multi-point conformal correlation functions organize the CFT data in
non-trivial functions of conformal cross ratios. These functions have,
generically, a complex analytic structure that does not follow from a
single exchange of a physical operator. Instead, it is often the case
that the intricate structure only emerges after summing
the contributions of an infinite number of
operators~\cite{Alday:2013cwa,Bercini:2020msp,Bercini:2021jti,Harris:2024nmr,Costa:2023wfz}.

Using the conformal bootstrap, we analyzed the five-point correlation
function of one Lagrangian and four lightest scalar operators, in
terms of the three-point functions of two leading-twist spinning
operators and the interaction Lagrangian. We computed these structure
constants for finite and large values of spin, connecting them with
anomalous dimensions~\eqref{CPTFinal}, null pentagon
Wilson loops~\eqref{eqCNullPent}, null square Wilson loops with
insertions~\eqref{eqFDerivation}, and the cusp anomalous
dimension~\eqref{eqCusptoB}.

In $\mathcal{N}=4$ SYM, there are several distinct integrability
frameworks developed to study the different observables listed above.
Three-point correlation functions are described by integrable hexagon
form factors~\cite{Basso:2015zoa}, null polygonal Wilson loops can be
constructed out of integrable pentagons~\cite{Basso:2013vsa}, and
anomalous dimensions can be computed via the quantum spectral
curve~\cite{Gromov:2013pga}. The sharp maps that we derived here
connect all these quantities and could be a great laboratory for
developing a unifying integrability description of $\mathcal{N}=4$~SYM.

It would be interesting to study the expectation value of
the square Wilson loop with other types of insertions using the techniques developed here. It should also be possible
and very interesting to generalize our analysis to other
physical observables, for example null square Wilson loops with two
operator insertion, or null pentagon Wilson loops with a single
operator insertion \cite{Chicherin:2022zxo}, and to connect these
quantities with conformal manifold
constraints~\cite{Behan:2017mwi,Hollands:2017chb} and integrability.

\section*{Acknowledgments}

We would like to thank Antonio Antunes, Pedro Vieira, Simon Ekhammar,
Nikolay Gromov and Gregory Korchemsky for illuminating discussions.
Centro de F\'{i}sica do Porto is partially funded by Funda\c{c}\~{a}o
para
a Ci\^{e}ncia e a Tecnologia (FCT) under Grant No.~UID04650-FCUP. The work of T.B.\ and C.B.\ was
funded by the Deutsche Forschungsgemeinschaft (DFG, German Research
Foundation) Grant No.~460391856.
T.B.\ and C.B.\ acknowledge support from DESY (Hamburg, Germany), a member of the Helmholtz Association HGF.
V.G.\ is supported by Simons Foundation under Grant No.~488637 (Simons collaboration
on the non-perturbative bootstrap) and Fundacao para a Ciencia e Tecnologia (FCT) under
Grant No.~CEECIND/03356/2022. B.F.\ is supported by the Simons
Foundation under Grant No.~488637
(Simons collaboration on the non-perturbative bootstrap) and by  Funda\c{c}\~{a}o para
a Ci\^{e}ncia e a Tecnologia, under the IDPASC doctoral program, under
Grant No.~PRT/BD/154692/2022. J.A.M.\ was supported by the Royal
Society under Grant No.~URF$\backslash$R1$\backslash$211417 and by the
European Research Council (ERC) under the European Union’s Horizon
2020 research and innovation
program 60 (Grant Agreement No.~865075) EXACTC.

\pdfbookmark[1]{\refname}{references}
\bibliography{references}

\begin{thebibliography}{40}%
\makeatletter
\providecommand \@ifxundefined [1]{%
 \@ifx{#1\undefined}
}%
\providecommand \@ifnum [1]{%
 \ifnum #1\expandafter \@firstoftwo
 \else \expandafter \@secondoftwo
 \fi
}%
\providecommand \@ifx [1]{%
 \ifx #1\expandafter \@firstoftwo
 \else \expandafter \@secondoftwo
 \fi
}%
\providecommand \natexlab [1]{#1}%
\providecommand \enquote  [1]{``#1''}%
\providecommand \bibnamefont  [1]{#1}%
\providecommand \bibfnamefont [1]{#1}%
\providecommand \citenamefont [1]{#1}%
\providecommand \href@noop [0]{\@secondoftwo}%
\providecommand \href [0]{\begingroup \@sanitize@url \@href}%
\providecommand \@href[1]{\@@startlink{#1}\@@href}%
\providecommand \@@href[1]{\endgroup#1\@@endlink}%
\providecommand \@sanitize@url [0]{\catcode `\\12\catcode `\$12\catcode
  `\&12\catcode `\#12\catcode `\^12\catcode `\_12\catcode `\%12\relax}%
\providecommand \@@startlink[1]{}%
\providecommand \@@endlink[0]{}%
\providecommand \url  [0]{\begingroup\@sanitize@url \@url }%
\providecommand \@url [1]{\endgroup\@href {#1}{\urlprefix }}%
\providecommand \urlprefix  [0]{URL }%
\providecommand \Eprint [0]{\href }%
\providecommand \doibase [0]{https://doi.org/}%
\providecommand \selectlanguage [0]{\@gobble}%
\providecommand \bibinfo  [0]{\@secondoftwo}%
\providecommand \bibfield  [0]{\@secondoftwo}%
\providecommand \translation [1]{[#1]}%
\providecommand \BibitemOpen [0]{}%
\providecommand \bibitemStop [0]{}%
\providecommand \bibitemNoStop [0]{.\EOS\space}%
\providecommand \EOS [0]{\spacefactor3000\relax}%
\providecommand \BibitemShut  [1]{\csname bibitem#1\endcsname}%
\let\auto@bib@innerbib\@empty
\bibitem [{\citenamefont {Bercini}\ \emph {et~al.}(2021)\citenamefont
  {Bercini}, \citenamefont {Gon\c{c}alves},\ and\ \citenamefont
  {Vieira}}]{Bercini:2020msp}%
  \BibitemOpen
  \bibfield  {author} {\bibinfo {author} {\bibfnamefont {C.}~\bibnamefont
  {Bercini}}, \bibinfo {author} {\bibfnamefont {V.}~\bibnamefont
  {Gon\c{c}alves}},\ and\ \bibinfo {author} {\bibfnamefont {P.}~\bibnamefont
  {Vieira}},\ }\bibfield  {title} {\bibinfo {title} {Light-cone bootstrap of
  higher point functions and {Wilson} loop duality},\ }\href
  {https://doi.org/10.1103/PhysRevLett.126.121603} {\bibfield  {journal}
  {\bibinfo  {journal} {Phys. Rev. Lett.}\ }\textbf {\bibinfo {volume} {126}},\
  \bibinfo {pages} {121603} (\bibinfo {year} {2021})},\ \Eprint
  {https://arxiv.org/abs/2008.10407} {arXiv:2008.10407 [hep-th]} \BibitemShut
  {NoStop}%
\bibitem [{\citenamefont {Bercini}\ \emph
  {et~al.}(2022{\natexlab{a}})\citenamefont {Bercini}, \citenamefont
  {Gon\c{c}alves}, \citenamefont {Homrich},\ and\ \citenamefont
  {Vieira}}]{Bercini:2021jti}%
  \BibitemOpen
  \bibfield  {author} {\bibinfo {author} {\bibfnamefont {C.}~\bibnamefont
  {Bercini}}, \bibinfo {author} {\bibfnamefont {V.}~\bibnamefont
  {Gon\c{c}alves}}, \bibinfo {author} {\bibfnamefont {A.}~\bibnamefont
  {Homrich}},\ and\ \bibinfo {author} {\bibfnamefont {P.}~\bibnamefont
  {Vieira}},\ }\bibfield  {title} {\bibinfo {title} {The {Wilson} loop
  \textemdash{} large spin ope dictionary},\ }\href
  {https://doi.org/10.1007/JHEP07(2022)079} {\bibfield  {journal} {\bibinfo
  {journal} {JHEP}\ }\textbf {\bibinfo {volume} {07}},\ \bibinfo {pages}
  {079}},\ \Eprint {https://arxiv.org/abs/2110.04364} {arXiv:2110.04364
  [hep-th]} \BibitemShut {NoStop}%
\bibitem [{\citenamefont {Antunes}\ \emph {et~al.}(2022)\citenamefont
  {Antunes}, \citenamefont {Costa}, \citenamefont {Gon\c{c}alves},\ and\
  \citenamefont {Boas}}]{Antunes:2021kmm}%
  \BibitemOpen
  \bibfield  {author} {\bibinfo {author} {\bibfnamefont {A.}~\bibnamefont
  {Antunes}}, \bibinfo {author} {\bibfnamefont {M.~S.}\ \bibnamefont {Costa}},
  \bibinfo {author} {\bibfnamefont {V.}~\bibnamefont {Gon\c{c}alves}},\ and\
  \bibinfo {author} {\bibfnamefont {J.~V.}\ \bibnamefont {Boas}},\ }\bibfield
  {title} {\bibinfo {title} {Lightcone bootstrap at higher points},\ }\href
  {https://doi.org/10.1007/JHEP03(2022)139} {\bibfield  {journal} {\bibinfo
  {journal} {JHEP}\ }\textbf {\bibinfo {volume} {03}},\ \bibinfo {pages}
  {139}},\ \Eprint {https://arxiv.org/abs/2111.05453} {arXiv:2111.05453
  [hep-th]} \BibitemShut {NoStop}%
\bibitem [{\citenamefont {Buric}\ and\ \citenamefont
  {Schomerus}(2023)}]{Buric:2022ucg}%
  \BibitemOpen
  \bibfield  {author} {\bibinfo {author} {\bibfnamefont {I.}~\bibnamefont
  {Buric}}\ and\ \bibinfo {author} {\bibfnamefont {V.}~\bibnamefont
  {Schomerus}},\ }\bibfield  {title} {\bibinfo {title} {Universal spinning
  {Casimir} equations and their solutions},\ }\href
  {https://doi.org/10.1007/JHEP03(2023)133} {\bibfield  {journal} {\bibinfo
  {journal} {JHEP}\ }\textbf {\bibinfo {volume} {03}},\ \bibinfo {pages}
  {133}},\ \Eprint {https://arxiv.org/abs/2211.14340} {arXiv:2211.14340
  [hep-th]} \BibitemShut {NoStop}%
\bibitem [{\citenamefont {Kaviraj}\ \emph {et~al.}(2023)\citenamefont
  {Kaviraj}, \citenamefont {Mann}, \citenamefont {Quintavalle},\ and\
  \citenamefont {Schomerus}}]{Kaviraj:2022wbw}%
  \BibitemOpen
  \bibfield  {author} {\bibinfo {author} {\bibfnamefont {A.}~\bibnamefont
  {Kaviraj}}, \bibinfo {author} {\bibfnamefont {J.~A.}\ \bibnamefont {Mann}},
  \bibinfo {author} {\bibfnamefont {L.}~\bibnamefont {Quintavalle}},\ and\
  \bibinfo {author} {\bibfnamefont {V.}~\bibnamefont {Schomerus}},\ }\bibfield
  {title} {\bibinfo {title} {Multipoint lightcone bootstrap from differential
  equations},\ }\href {https://doi.org/10.1007/JHEP08(2023)011} {\bibfield
  {journal} {\bibinfo  {journal} {JHEP}\ }\textbf {\bibinfo {volume} {08}},\
  \bibinfo {pages} {011}},\ \Eprint {https://arxiv.org/abs/2212.10578}
  {arXiv:2212.10578 [hep-th]} \BibitemShut {NoStop}%
\bibitem [{\citenamefont {Poland}\ \emph {et~al.}(2023)\citenamefont {Poland},
  \citenamefont {Prilepina},\ and\ \citenamefont {Tadi\'c}}]{Poland:2023vpn}%
  \BibitemOpen
  \bibfield  {author} {\bibinfo {author} {\bibfnamefont {D.}~\bibnamefont
  {Poland}}, \bibinfo {author} {\bibfnamefont {V.}~\bibnamefont {Prilepina}},\
  and\ \bibinfo {author} {\bibfnamefont {P.}~\bibnamefont {Tadi\'c}},\
  }\bibfield  {title} {\bibinfo {title} {The five-point bootstrap},\ }\href
  {https://doi.org/10.1007/JHEP10(2023)153} {\bibfield  {journal} {\bibinfo
  {journal} {JHEP}\ }\textbf {\bibinfo {volume} {10}},\ \bibinfo {pages}
  {153}},\ \Eprint {https://arxiv.org/abs/2305.08914} {arXiv:2305.08914
  [hep-th]} \BibitemShut {NoStop}%
\bibitem [{\citenamefont {Antunes}\ \emph {et~al.}(2024)\citenamefont
  {Antunes}, \citenamefont {Harris}, \citenamefont {Kaviraj},\ and\
  \citenamefont {Schomerus}}]{Antunes:2023kyz}%
  \BibitemOpen
  \bibfield  {author} {\bibinfo {author} {\bibfnamefont {A.}~\bibnamefont
  {Antunes}}, \bibinfo {author} {\bibfnamefont {S.}~\bibnamefont {Harris}},
  \bibinfo {author} {\bibfnamefont {A.}~\bibnamefont {Kaviraj}},\ and\ \bibinfo
  {author} {\bibfnamefont {V.}~\bibnamefont {Schomerus}},\ }\bibfield  {title}
  {\bibinfo {title} {Lining up a positive semi-definite six-point bootstrap},\
  }\href {https://doi.org/10.1007/JHEP06(2024)058} {\bibfield  {journal}
  {\bibinfo  {journal} {JHEP}\ }\textbf {\bibinfo {volume} {06}},\ \bibinfo
  {pages} {058}},\ \Eprint {https://arxiv.org/abs/2312.11660} {arXiv:2312.11660
  [hep-th]} \BibitemShut {NoStop}%
\bibitem [{\citenamefont {Poland}\ \emph {et~al.}(2024)\citenamefont {Poland},
  \citenamefont {Prilepina},\ and\ \citenamefont {Tadi\'c}}]{Poland:2023bny}%
  \BibitemOpen
  \bibfield  {author} {\bibinfo {author} {\bibfnamefont {D.}~\bibnamefont
  {Poland}}, \bibinfo {author} {\bibfnamefont {V.}~\bibnamefont {Prilepina}},\
  and\ \bibinfo {author} {\bibfnamefont {P.}~\bibnamefont {Tadi\'c}},\
  }\bibfield  {title} {\bibinfo {title} {Improving the five-point bootstrap},\
  }\href {https://doi.org/10.1007/JHEP05(2024)299} {\bibfield  {journal}
  {\bibinfo  {journal} {JHEP}\ }\textbf {\bibinfo {volume} {05}},\ \bibinfo
  {pages} {299}},\ \Eprint {https://arxiv.org/abs/2312.13344} {arXiv:2312.13344
  [hep-th]} \BibitemShut {NoStop}%
\bibitem [{\citenamefont {Harris}\ \emph {et~al.}(2024)\citenamefont {Harris},
  \citenamefont {Kaviraj}, \citenamefont {Mann}, \citenamefont {Quintavalle},\
  and\ \citenamefont {Schomerus}}]{Harris:2024nmr}%
  \BibitemOpen
  \bibfield  {author} {\bibinfo {author} {\bibfnamefont {S.}~\bibnamefont
  {Harris}}, \bibinfo {author} {\bibfnamefont {A.}~\bibnamefont {Kaviraj}},
  \bibinfo {author} {\bibfnamefont {J.~A.}\ \bibnamefont {Mann}}, \bibinfo
  {author} {\bibfnamefont {L.}~\bibnamefont {Quintavalle}},\ and\ \bibinfo
  {author} {\bibfnamefont {V.}~\bibnamefont {Schomerus}},\ }\bibfield  {title}
  {\bibinfo {title} {Comb channel lightcone bootstrap: triple-twist anomalous
  dimensions},\ }\href {https://doi.org/10.1007/JHEP08(2024)122} {\bibfield
  {journal} {\bibinfo  {journal} {JHEP}\ }\textbf {\bibinfo {volume} {08}},\
  \bibinfo {pages} {122}},\ \Eprint {https://arxiv.org/abs/2401.10986}
  {arXiv:2401.10986 [hep-th]} \BibitemShut {NoStop}%
\bibitem [{\citenamefont {Alday}\ and\ \citenamefont
  {Bissi}(2013)}]{Alday:2013cwa}%
  \BibitemOpen
  \bibfield  {author} {\bibinfo {author} {\bibfnamefont {L.~F.}\ \bibnamefont
  {Alday}}\ and\ \bibinfo {author} {\bibfnamefont {A.}~\bibnamefont {Bissi}},\
  }\bibfield  {title} {\bibinfo {title} {Higher-spin correlators},\ }\href
  {https://doi.org/10.1007/JHEP10(2013)202} {\bibfield  {journal} {\bibinfo
  {journal} {JHEP}\ }\textbf {\bibinfo {volume} {10}},\ \bibinfo {pages}
  {202}},\ \Eprint {https://arxiv.org/abs/1305.4604} {arXiv:1305.4604 [hep-th]}
  \BibitemShut {NoStop}%
\bibitem [{\citenamefont {Eden}\ \emph {et~al.}(2012)\citenamefont {Eden},
  \citenamefont {Heslop}, \citenamefont {Korchemsky},\ and\ \citenamefont
  {Sokatchev}}]{Eden:2011we}%
  \BibitemOpen
  \bibfield  {author} {\bibinfo {author} {\bibfnamefont {B.}~\bibnamefont
  {Eden}}, \bibinfo {author} {\bibfnamefont {P.}~\bibnamefont {Heslop}},
  \bibinfo {author} {\bibfnamefont {G.~P.}\ \bibnamefont {Korchemsky}},\ and\
  \bibinfo {author} {\bibfnamefont {E.}~\bibnamefont {Sokatchev}},\ }\bibfield
  {title} {\bibinfo {title} {Hidden symmetry of four-point correlation
  functions and amplitudes in {$\mathcal{N}=\mathord{}$4} {SYM}},\ }\href
  {https://doi.org/10.1016/j.nuclphysb.2012.04.007} {\bibfield  {journal}
  {\bibinfo  {journal} {Nucl. Phys. B}\ }\textbf {\bibinfo {volume} {862}},\
  \bibinfo {pages} {193} (\bibinfo {year} {2012})},\ \Eprint
  {https://arxiv.org/abs/1108.3557} {arXiv:1108.3557 [hep-th]} \BibitemShut
  {NoStop}%
\bibitem [{\citenamefont {Alday}\ \emph
  {et~al.}(2011{\natexlab{a}})\citenamefont {Alday}, \citenamefont
  {Buchbinder},\ and\ \citenamefont {Tseytlin}}]{Alday:2011ga}%
  \BibitemOpen
  \bibfield  {author} {\bibinfo {author} {\bibfnamefont {L.~F.}\ \bibnamefont
  {Alday}}, \bibinfo {author} {\bibfnamefont {E.~I.}\ \bibnamefont
  {Buchbinder}},\ and\ \bibinfo {author} {\bibfnamefont {A.~A.}\ \bibnamefont
  {Tseytlin}},\ }\bibfield  {title} {\bibinfo {title} {Correlation function of
  null polygonal {Wilson} loops with local operators},\ }\href
  {https://doi.org/10.1007/JHEP09(2011)034} {\bibfield  {journal} {\bibinfo
  {journal} {JHEP}\ }\textbf {\bibinfo {volume} {1109}},\ \bibinfo {pages}
  {034}},\ \Eprint {https://arxiv.org/abs/1107.5702} {arXiv:1107.5702 [hep-th]}
  \BibitemShut {NoStop}%
\bibitem [{\citenamefont {Alday}\ \emph
  {et~al.}(2013{\natexlab{a}})\citenamefont {Alday}, \citenamefont {Heslop},\
  and\ \citenamefont {Sikorowski}}]{Alday:2012hy}%
  \BibitemOpen
  \bibfield  {author} {\bibinfo {author} {\bibfnamefont {L.~F.}\ \bibnamefont
  {Alday}}, \bibinfo {author} {\bibfnamefont {P.}~\bibnamefont {Heslop}},\ and\
  \bibinfo {author} {\bibfnamefont {J.}~\bibnamefont {Sikorowski}},\ }\bibfield
   {title} {\bibinfo {title} {Perturbative correlation functions of null
  {Wilson} loops and local operators},\ }\href
  {https://doi.org/10.1007/JHEP03(2013)074} {\bibfield  {journal} {\bibinfo
  {journal} {JHEP}\ }\textbf {\bibinfo {volume} {03}},\ \bibinfo {pages}
  {074}},\ \Eprint {https://arxiv.org/abs/1207.4316} {arXiv:1207.4316 [hep-th]}
  \BibitemShut {NoStop}%
\bibitem [{\citenamefont {Alday}\ \emph
  {et~al.}(2013{\natexlab{b}})\citenamefont {Alday}, \citenamefont {Henn},\
  and\ \citenamefont {Sikorowski}}]{Alday:2013ip}%
  \BibitemOpen
  \bibfield  {author} {\bibinfo {author} {\bibfnamefont {L.~F.}\ \bibnamefont
  {Alday}}, \bibinfo {author} {\bibfnamefont {J.~M.}\ \bibnamefont {Henn}},\
  and\ \bibinfo {author} {\bibfnamefont {J.}~\bibnamefont {Sikorowski}},\
  }\bibfield  {title} {\bibinfo {title} {Higher loop mixed correlators in
  {$\mathcal{N}=\mathord{}$4} {SYM}},\ }\href
  {https://doi.org/10.1007/JHEP03(2013)058} {\bibfield  {journal} {\bibinfo
  {journal} {JHEP}\ }\textbf {\bibinfo {volume} {03}},\ \bibinfo {pages}
  {058}},\ \Eprint {https://arxiv.org/abs/1301.0149} {arXiv:1301.0149 [hep-th]}
  \BibitemShut {NoStop}%
\bibitem [{\citenamefont {Henn}\ \emph {et~al.}(2020)\citenamefont {Henn},
  \citenamefont {Korchemsky},\ and\ \citenamefont
  {Mistlberger}}]{Henn:2019swt}%
  \BibitemOpen
  \bibfield  {author} {\bibinfo {author} {\bibfnamefont {J.~M.}\ \bibnamefont
  {Henn}}, \bibinfo {author} {\bibfnamefont {G.~P.}\ \bibnamefont
  {Korchemsky}},\ and\ \bibinfo {author} {\bibfnamefont {B.}~\bibnamefont
  {Mistlberger}},\ }\bibfield  {title} {\bibinfo {title} {The full four-loop
  cusp anomalous dimension in {$\mathcal{N}=\mathord{}$4} super {Yang}--{Mills}
  and qcd},\ }\href {https://doi.org/10.1007/JHEP04(2020)018} {\bibfield
  {journal} {\bibinfo  {journal} {JHEP}\ }\textbf {\bibinfo {volume} {04}},\
  \bibinfo {pages} {018}},\ \Eprint {https://arxiv.org/abs/1911.10174}
  {arXiv:1911.10174 [hep-th]} \BibitemShut {NoStop}%
\bibitem [{\citenamefont {Ferrara}\ \emph {et~al.}(1974)\citenamefont
  {Ferrara}, \citenamefont {Grillo}, \citenamefont {Gatto},\ and\ \citenamefont
  {Parisi}}]{Ferrara:1974nf}%
  \BibitemOpen
  \bibfield  {author} {\bibinfo {author} {\bibfnamefont {S.}~\bibnamefont
  {Ferrara}}, \bibinfo {author} {\bibfnamefont {A.~F.}\ \bibnamefont {Grillo}},
  \bibinfo {author} {\bibfnamefont {R.}~\bibnamefont {Gatto}},\ and\ \bibinfo
  {author} {\bibfnamefont {G.}~\bibnamefont {Parisi}},\ }\bibfield  {title}
  {\bibinfo {title} {Analyticity properties and asymptotic expansions of
  conformal covariant green's functions},\ }\href
  {https://doi.org/10.1007/BF02813413} {\bibfield  {journal} {\bibinfo
  {journal} {Nuovo Cim. A}\ }\textbf {\bibinfo {volume} {19}},\ \bibinfo
  {pages} {667} (\bibinfo {year} {1974})}\BibitemShut {NoStop}%
\bibitem [{\citenamefont {Costa}\ \emph {et~al.}(2011)\citenamefont {Costa},
  \citenamefont {Penedones}, \citenamefont {Poland},\ and\ \citenamefont
  {Rychkov}}]{Costa:2011dw}%
  \BibitemOpen
  \bibfield  {author} {\bibinfo {author} {\bibfnamefont {M.~S.}\ \bibnamefont
  {Costa}}, \bibinfo {author} {\bibfnamefont {J.}~\bibnamefont {Penedones}},
  \bibinfo {author} {\bibfnamefont {D.}~\bibnamefont {Poland}},\ and\ \bibinfo
  {author} {\bibfnamefont {S.}~\bibnamefont {Rychkov}},\ }\bibfield  {title}
  {\bibinfo {title} {Spinning conformal blocks},\ }\href
  {https://doi.org/10.1007/JHEP11(2011)154} {\bibfield  {journal} {\bibinfo
  {journal} {JHEP}\ }\textbf {\bibinfo {volume} {11}},\ \bibinfo {pages}
  {154}},\ \Eprint {https://arxiv.org/abs/1109.6321} {arXiv:1109.6321 [hep-th]}
  \BibitemShut {NoStop}%
\bibitem [{\citenamefont {Basso}\ \emph {et~al.}(2015)\citenamefont {Basso},
  \citenamefont {Komatsu},\ and\ \citenamefont {Vieira}}]{Basso:2015zoa}%
  \BibitemOpen
  \bibfield  {author} {\bibinfo {author} {\bibfnamefont {B.}~\bibnamefont
  {Basso}}, \bibinfo {author} {\bibfnamefont {S.}~\bibnamefont {Komatsu}},\
  and\ \bibinfo {author} {\bibfnamefont {P.}~\bibnamefont {Vieira}},\
  }\href@noop {} {\bibinfo {title} {Structure constants and integrable
  bootstrap in planar {$\mathcal{N}=\mathord{}$4} {SYM} theory}} (\bibinfo
  {year} {2015}),\ \Eprint {https://arxiv.org/abs/1505.06745} {arXiv:1505.06745
  [hep-th]} \BibitemShut {NoStop}%
\bibitem [{\citenamefont {Bercini}\ \emph
  {et~al.}(2022{\natexlab{b}})\citenamefont {Bercini}, \citenamefont
  {Gon\c{c}alves}, \citenamefont {Homrich},\ and\ \citenamefont
  {Vieira}}]{Bercini:2022gvs}%
  \BibitemOpen
  \bibfield  {author} {\bibinfo {author} {\bibfnamefont {C.}~\bibnamefont
  {Bercini}}, \bibinfo {author} {\bibfnamefont {V.}~\bibnamefont
  {Gon\c{c}alves}}, \bibinfo {author} {\bibfnamefont {A.}~\bibnamefont
  {Homrich}},\ and\ \bibinfo {author} {\bibfnamefont {P.}~\bibnamefont
  {Vieira}},\ }\href@noop {} {\bibinfo {title} {Spinning hexagons}} (\bibinfo
  {year} {2022}{\natexlab{b}}),\ \Eprint {https://arxiv.org/abs/2207.08931}
  {arXiv:2207.08931 [hep-th]} \BibitemShut {NoStop}%
\bibitem [{\citenamefont {Eden}\ \emph {et~al.}(2024)\citenamefont {Eden},
  \citenamefont {Gottwald}, \citenamefont {le~Plat},\ and\ \citenamefont
  {Scherdin}}]{Eden:2023gso}%
  \BibitemOpen
  \bibfield  {author} {\bibinfo {author} {\bibfnamefont {B.}~\bibnamefont
  {Eden}}, \bibinfo {author} {\bibfnamefont {M.}~\bibnamefont {Gottwald}},
  \bibinfo {author} {\bibfnamefont {D.}~\bibnamefont {le~Plat}},\ and\ \bibinfo
  {author} {\bibfnamefont {T.}~\bibnamefont {Scherdin}},\ }\bibfield  {title}
  {\bibinfo {title} {Anomalous dimensions from the {$\mathcal{N}=\mathord{}$4}
  supersymmetric {Yang}--{Mills} hexagon},\ }\href
  {https://doi.org/10.1103/PhysRevLett.132.161605} {\bibfield  {journal}
  {\bibinfo  {journal} {Phys. Rev. Lett.}\ }\textbf {\bibinfo {volume} {132}},\
  \bibinfo {pages} {161605} (\bibinfo {year} {2024})},\ \Eprint
  {https://arxiv.org/abs/2310.04392} {arXiv:2310.04392 [hep-th]} \BibitemShut
  {NoStop}%
\bibitem [{\citenamefont {Bercini}\ \emph {et~al.}(2024)\citenamefont
  {Bercini}, \citenamefont {Fernandes},\ and\ \citenamefont
  {Gon\c{c}alves}}]{Bercini:2024pya}%
  \BibitemOpen
  \bibfield  {author} {\bibinfo {author} {\bibfnamefont {C.}~\bibnamefont
  {Bercini}}, \bibinfo {author} {\bibfnamefont {B.}~\bibnamefont {Fernandes}},\
  and\ \bibinfo {author} {\bibfnamefont {V.}~\bibnamefont {Gon\c{c}alves}},\
  }\bibfield  {title} {\bibinfo {title} {Two-loop five-point integrals: light,
  heavy and large-spin correlators},\ }\href
  {https://doi.org/10.1007/JHEP10(2024)242} {\bibfield  {journal} {\bibinfo
  {journal} {JHEP}\ }\textbf {\bibinfo {volume} {10}},\ \bibinfo {pages}
  {242}},\ \Eprint {https://arxiv.org/abs/2401.06099} {arXiv:2401.06099
  [hep-th]} \BibitemShut {NoStop}%
\bibitem [{\citenamefont {Zamolodchikov}(1987)}]{Zamolodchikov:1987ti}%
  \BibitemOpen
  \bibfield  {author} {\bibinfo {author} {\bibfnamefont {A.~B.}\ \bibnamefont
  {Zamolodchikov}},\ }\bibfield  {title} {\bibinfo {title} {Renormalization
  group and perturbation theory near fixed points in two-dimensional field
  theory},\ }\href@noop {} {\bibfield  {journal} {\bibinfo  {journal} {Sov. J.
  Nucl. Phys.}\ }\textbf {\bibinfo {volume} {46}},\ \bibinfo {pages} {1090}
  (\bibinfo {year} {1987})}\BibitemShut {NoStop}%
\bibitem [{\citenamefont {Sen}\ and\ \citenamefont
  {Tachikawa}(2017)}]{Sen:2017gfr}%
  \BibitemOpen
  \bibfield  {author} {\bibinfo {author} {\bibfnamefont {K.}~\bibnamefont
  {Sen}}\ and\ \bibinfo {author} {\bibfnamefont {Y.}~\bibnamefont
  {Tachikawa}},\ }\href@noop {} {\bibinfo {title} {First-order conformal
  perturbation theory by marginal operators}} (\bibinfo {year} {2017}),\
  \Eprint {https://arxiv.org/abs/1711.05947} {arXiv:1711.05947 [hep-th]}
  \BibitemShut {NoStop}%
\bibitem [{\citenamefont {Korchemsky}(1989)}]{Korchemsky:1988si}%
  \BibitemOpen
  \bibfield  {author} {\bibinfo {author} {\bibfnamefont {G.~P.}\ \bibnamefont
  {Korchemsky}},\ }\bibfield  {title} {\bibinfo {title} {Asymptotics of the
  altarelli-parisi-lipatov evolution kernels of parton distributions},\ }\href
  {https://doi.org/10.1142/S0217732389001453} {\bibfield  {journal} {\bibinfo
  {journal} {Mod. Phys. Lett. A}\ }\textbf {\bibinfo {volume} {4}},\ \bibinfo
  {pages} {1257} (\bibinfo {year} {1989})}\BibitemShut {NoStop}%
\bibitem [{\citenamefont {Gubser}\ \emph {et~al.}(2002)\citenamefont {Gubser},
  \citenamefont {Klebanov},\ and\ \citenamefont {Polyakov}}]{Gubser:2002tv}%
  \BibitemOpen
  \bibfield  {author} {\bibinfo {author} {\bibfnamefont {S.~S.}\ \bibnamefont
  {Gubser}}, \bibinfo {author} {\bibfnamefont {I.~R.}\ \bibnamefont
  {Klebanov}},\ and\ \bibinfo {author} {\bibfnamefont {A.~M.}\ \bibnamefont
  {Polyakov}},\ }\bibfield  {title} {\bibinfo {title} {A semi-classical limit
  of the gauge/string correspondence},\ }\href
  {https://doi.org/10.1016/S0550-3213(02)00373-5} {\bibfield  {journal}
  {\bibinfo  {journal} {Nucl. Phys.}\ }\textbf {\bibinfo {volume} {B636}},\
  \bibinfo {pages} {99} (\bibinfo {year} {2002})},\ \Eprint
  {https://arxiv.org/abs/hep-th/0204051} {arXiv:hep-th/0204051} \BibitemShut
  {NoStop}%
\bibitem [{\citenamefont {Alday}\ \emph {et~al.}(2015)\citenamefont {Alday},
  \citenamefont {Bissi},\ and\ \citenamefont {Lukowski}}]{Alday:2015eya}%
  \BibitemOpen
  \bibfield  {author} {\bibinfo {author} {\bibfnamefont {L.~F.}\ \bibnamefont
  {Alday}}, \bibinfo {author} {\bibfnamefont {A.}~\bibnamefont {Bissi}},\ and\
  \bibinfo {author} {\bibfnamefont {T.}~\bibnamefont {Lukowski}},\ }\bibfield
  {title} {\bibinfo {title} {Large spin systematics in cft},\ }\href
  {https://doi.org/10.1007/JHEP11(2015)101} {\bibfield  {journal} {\bibinfo
  {journal} {JHEP}\ }\textbf {\bibinfo {volume} {11}},\ \bibinfo {pages}
  {101}},\ \Eprint {https://arxiv.org/abs/1502.07707} {arXiv:1502.07707
  [hep-th]} \BibitemShut {NoStop}%
\bibitem [{\citenamefont {Alday}\ and\ \citenamefont
  {Zhiboedov}(2017)}]{Alday:2015ewa}%
  \BibitemOpen
  \bibfield  {author} {\bibinfo {author} {\bibfnamefont {L.~F.}\ \bibnamefont
  {Alday}}\ and\ \bibinfo {author} {\bibfnamefont {A.}~\bibnamefont
  {Zhiboedov}},\ }\bibfield  {title} {\bibinfo {title} {An algebraic approach
  to the analytic bootstrap},\ }\href {https://doi.org/10.1007/JHEP04(2017)157}
  {\bibfield  {journal} {\bibinfo  {journal} {JHEP}\ }\textbf {\bibinfo
  {volume} {04}},\ \bibinfo {pages} {157}},\ \Eprint
  {https://arxiv.org/abs/1510.08091} {arXiv:1510.08091 [hep-th]} \BibitemShut
  {NoStop}%
\bibitem [{\citenamefont {Simmons-Duffin}(2017)}]{Simmons-Duffin:2016wlq}%
  \BibitemOpen
  \bibfield  {author} {\bibinfo {author} {\bibfnamefont {D.}~\bibnamefont
  {Simmons-Duffin}},\ }\bibfield  {title} {\bibinfo {title} {The lightcone
  bootstrap and the spectrum of the 3d ising cft},\ }\href
  {https://doi.org/10.1007/JHEP03(2017)086} {\bibfield  {journal} {\bibinfo
  {journal} {JHEP}\ }\textbf {\bibinfo {volume} {03}},\ \bibinfo {pages}
  {086}},\ \Eprint {https://arxiv.org/abs/1612.08471} {arXiv:1612.08471
  [hep-th]} \BibitemShut {NoStop}%
\bibitem [{\citenamefont {Remiddi}\ and\ \citenamefont
  {Vermaseren}(2000)}]{Remiddi:1999ew}%
  \BibitemOpen
  \bibfield  {author} {\bibinfo {author} {\bibfnamefont {E.}~\bibnamefont
  {Remiddi}}\ and\ \bibinfo {author} {\bibfnamefont {J.~A.~M.}\ \bibnamefont
  {Vermaseren}},\ }\bibfield  {title} {\bibinfo {title} {Harmonic
  polylogarithms},\ }\href {https://doi.org/10.1142/S0217751X00000367}
  {\bibfield  {journal} {\bibinfo  {journal} {Int. J. Mod. Phys. A}\ }\textbf
  {\bibinfo {volume} {15}},\ \bibinfo {pages} {725} (\bibinfo {year} {2000})},\
  \Eprint {https://arxiv.org/abs/hep-ph/9905237} {arXiv:hep-ph/9905237}
  \BibitemShut {NoStop}%
\bibitem [{\citenamefont {Maitre}(2006)}]{Maitre:2005uu}%
  \BibitemOpen
  \bibfield  {author} {\bibinfo {author} {\bibfnamefont {D.}~\bibnamefont
  {Maitre}},\ }\bibfield  {title} {\bibinfo {title} {Hpl, a mathematica
  implementation of the harmonic polylogarithms},\ }\href
  {https://doi.org/10.1016/j.cpc.2005.10.008} {\bibfield  {journal} {\bibinfo
  {journal} {Comput. Phys. Commun.}\ }\textbf {\bibinfo {volume} {174}},\
  \bibinfo {pages} {222} (\bibinfo {year} {2006})},\ \Eprint
  {https://arxiv.org/abs/hep-ph/0507152} {arXiv:hep-ph/0507152} \BibitemShut
  {NoStop}%
\bibitem [{\citenamefont {Alday}\ \emph
  {et~al.}(2011{\natexlab{b}})\citenamefont {Alday}, \citenamefont {Eden},
  \citenamefont {Korchemsky}, \citenamefont {Maldacena},\ and\ \citenamefont
  {Sokatchev}}]{Alday:2010zy}%
  \BibitemOpen
  \bibfield  {author} {\bibinfo {author} {\bibfnamefont {L.~F.}\ \bibnamefont
  {Alday}}, \bibinfo {author} {\bibfnamefont {B.}~\bibnamefont {Eden}},
  \bibinfo {author} {\bibfnamefont {G.~P.}\ \bibnamefont {Korchemsky}},
  \bibinfo {author} {\bibfnamefont {J.}~\bibnamefont {Maldacena}},\ and\
  \bibinfo {author} {\bibfnamefont {E.}~\bibnamefont {Sokatchev}},\ }\bibfield
  {title} {\bibinfo {title} {From correlation functions to {Wilson} loops},\
  }\href {https://doi.org/10.1007/JHEP09(2011)123} {\bibfield  {journal}
  {\bibinfo  {journal} {JHEP}\ }\textbf {\bibinfo {volume} {1109}},\ \bibinfo
  {pages} {123}},\ \Eprint {https://arxiv.org/abs/1007.3243} {arXiv:1007.3243
  [hep-th]} \BibitemShut {NoStop}%
\bibitem [{\citenamefont {Eden}\ \emph {et~al.}(2011)\citenamefont {Eden},
  \citenamefont {Korchemsky},\ and\ \citenamefont {Sokatchev}}]{Eden:2010zz}%
  \BibitemOpen
  \bibfield  {author} {\bibinfo {author} {\bibfnamefont {B.}~\bibnamefont
  {Eden}}, \bibinfo {author} {\bibfnamefont {G.~P.}\ \bibnamefont
  {Korchemsky}},\ and\ \bibinfo {author} {\bibfnamefont {E.}~\bibnamefont
  {Sokatchev}},\ }\bibfield  {title} {\bibinfo {title} {From correlation
  functions to scattering amplitudes},\ }\href
  {https://doi.org/10.1007/JHEP12(2011)002} {\bibfield  {journal} {\bibinfo
  {journal} {JHEP}\ }\textbf {\bibinfo {volume} {1112}},\ \bibinfo {pages}
  {002}},\ \Eprint {https://arxiv.org/abs/1007.3246} {arXiv:1007.3246 [hep-th]}
  \BibitemShut {NoStop}%
\bibitem [{\citenamefont {Eden}\ \emph
  {et~al.}(2013{\natexlab{a}})\citenamefont {Eden}, \citenamefont {Heslop},
  \citenamefont {Korchemsky},\ and\ \citenamefont {Sokatchev}}]{Eden:2011yp}%
  \BibitemOpen
  \bibfield  {author} {\bibinfo {author} {\bibfnamefont {B.}~\bibnamefont
  {Eden}}, \bibinfo {author} {\bibfnamefont {P.}~\bibnamefont {Heslop}},
  \bibinfo {author} {\bibfnamefont {G.~P.}\ \bibnamefont {Korchemsky}},\ and\
  \bibinfo {author} {\bibfnamefont {E.}~\bibnamefont {Sokatchev}},\ }\bibfield
  {title} {\bibinfo {title} {The super-correlator/super-amplitude duality: Part
  i},\ }\href {https://doi.org/10.1016/j.nuclphysb.2012.12.015} {\bibfield
  {journal} {\bibinfo  {journal} {Nucl. Phys.}\ }\textbf {\bibinfo {volume}
  {B869}},\ \bibinfo {pages} {329} (\bibinfo {year} {2013}{\natexlab{a}})},\
  \Eprint {https://arxiv.org/abs/1103.3714} {arXiv:1103.3714 [hep-th]}
  \BibitemShut {NoStop}%
\bibitem [{\citenamefont {Eden}\ \emph
  {et~al.}(2013{\natexlab{b}})\citenamefont {Eden}, \citenamefont {Heslop},
  \citenamefont {Korchemsky},\ and\ \citenamefont {Sokatchev}}]{Eden:2011ku}%
  \BibitemOpen
  \bibfield  {author} {\bibinfo {author} {\bibfnamefont {B.}~\bibnamefont
  {Eden}}, \bibinfo {author} {\bibfnamefont {P.}~\bibnamefont {Heslop}},
  \bibinfo {author} {\bibfnamefont {G.~P.}\ \bibnamefont {Korchemsky}},\ and\
  \bibinfo {author} {\bibfnamefont {E.}~\bibnamefont {Sokatchev}},\ }\bibfield
  {title} {\bibinfo {title} {The super-correlator/super-amplitude duality: Part
  {II}},\ }\href {https://doi.org/10.1016/j.nuclphysb.2012.12.014} {\bibfield
  {journal} {\bibinfo  {journal} {Nucl. Phys.}\ }\textbf {\bibinfo {volume}
  {B869}},\ \bibinfo {pages} {378} (\bibinfo {year} {2013}{\natexlab{b}})},\
  \Eprint {https://arxiv.org/abs/1103.4353} {arXiv:1103.4353 [hep-th]}
  \BibitemShut {NoStop}%
\bibitem [{\citenamefont {Costa}\ \emph {et~al.}(2023)\citenamefont {Costa},
  \citenamefont {Gon\c{c}alves}, \citenamefont {Salgarkar},\ and\ \citenamefont
  {Vilas~Boas}}]{Costa:2023wfz}%
  \BibitemOpen
  \bibfield  {author} {\bibinfo {author} {\bibfnamefont {M.~S.}\ \bibnamefont
  {Costa}}, \bibinfo {author} {\bibfnamefont {V.}~\bibnamefont
  {Gon\c{c}alves}}, \bibinfo {author} {\bibfnamefont {A.}~\bibnamefont
  {Salgarkar}},\ and\ \bibinfo {author} {\bibfnamefont {J.}~\bibnamefont
  {Vilas~Boas}},\ }\bibfield  {title} {\bibinfo {title} {Conformal
  multi-{Regge} theory},\ }\href {https://doi.org/10.1007/JHEP09(2023)155}
  {\bibfield  {journal} {\bibinfo  {journal} {JHEP}\ }\textbf {\bibinfo
  {volume} {09}},\ \bibinfo {pages} {155}},\ \Eprint
  {https://arxiv.org/abs/2305.10394} {arXiv:2305.10394 [hep-th]} \BibitemShut
  {NoStop}%
\bibitem [{\citenamefont {Basso}\ \emph {et~al.}(2013)\citenamefont {Basso},
  \citenamefont {Sever},\ and\ \citenamefont {Vieira}}]{Basso:2013vsa}%
  \BibitemOpen
  \bibfield  {author} {\bibinfo {author} {\bibfnamefont {B.}~\bibnamefont
  {Basso}}, \bibinfo {author} {\bibfnamefont {A.}~\bibnamefont {Sever}},\ and\
  \bibinfo {author} {\bibfnamefont {P.}~\bibnamefont {Vieira}},\ }\bibfield
  {title} {\bibinfo {title} {Spacetime and flux tube {S}-matrices at finite
  coupling for {$\mathcal{N}=\mathord{}$4} supersymmetric {Yang}--{Mills}
  theory},\ }\href {https://doi.org/10.1103/PhysRevLett.111.091602} {\bibfield
  {journal} {\bibinfo  {journal} {Phys. Rev. Lett.}\ }\textbf {\bibinfo
  {volume} {111}},\ \bibinfo {pages} {091602} (\bibinfo {year} {2013})},\
  \Eprint {https://arxiv.org/abs/1303.1396} {arXiv:1303.1396 [hep-th]}
  \BibitemShut {NoStop}%
\bibitem [{\citenamefont {Gromov}\ \emph {et~al.}(2014)\citenamefont {Gromov},
  \citenamefont {Kazakov}, \citenamefont {Leurent},\ and\ \citenamefont
  {Volin}}]{Gromov:2013pga}%
  \BibitemOpen
  \bibfield  {author} {\bibinfo {author} {\bibfnamefont {N.}~\bibnamefont
  {Gromov}}, \bibinfo {author} {\bibfnamefont {V.}~\bibnamefont {Kazakov}},
  \bibinfo {author} {\bibfnamefont {S.}~\bibnamefont {Leurent}},\ and\ \bibinfo
  {author} {\bibfnamefont {D.}~\bibnamefont {Volin}},\ }\bibfield  {title}
  {\bibinfo {title} {Quantum spectral curve for planar
  {$\mathcal{N}=\mathord{}$4} super-{Yang}--{Mills} theory},\ }\href
  {https://doi.org/10.1103/PhysRevLett.112.011602} {\bibfield  {journal}
  {\bibinfo  {journal} {Phys. Rev. Lett.}\ }\textbf {\bibinfo {volume} {112}},\
  \bibinfo {pages} {011602} (\bibinfo {year} {2014})},\ \Eprint
  {https://arxiv.org/abs/1305.1939} {arXiv:1305.1939 [hep-th]} \BibitemShut
  {NoStop}%
\bibitem [{\citenamefont {Chicherin}\ and\ \citenamefont
  {Henn}(2022)}]{Chicherin:2022zxo}%
  \BibitemOpen
  \bibfield  {author} {\bibinfo {author} {\bibfnamefont {D.}~\bibnamefont
  {Chicherin}}\ and\ \bibinfo {author} {\bibfnamefont {J.}~\bibnamefont
  {Henn}},\ }\bibfield  {title} {\bibinfo {title} {Pentagon {Wilson} loop with
  lagrangian insertion at two loops in {$\mathcal{N}=\mathord{}$4} super
  {Yang}--{Mills} theory},\ }\href {https://doi.org/10.1007/JHEP07(2022)038}
  {\bibfield  {journal} {\bibinfo  {journal} {JHEP}\ }\textbf {\bibinfo
  {volume} {07}},\ \bibinfo {pages} {038}},\ \Eprint
  {https://arxiv.org/abs/2204.00329} {arXiv:2204.00329 [hep-th]} \BibitemShut
  {NoStop}%
\bibitem [{\citenamefont {Behan}(2018)}]{Behan:2017mwi}%
  \BibitemOpen
  \bibfield  {author} {\bibinfo {author} {\bibfnamefont {C.}~\bibnamefont
  {Behan}},\ }\bibfield  {title} {\bibinfo {title} {Conformal manifolds: Odes
  from opes},\ }\href {https://doi.org/10.1007/JHEP03(2018)127} {\bibfield
  {journal} {\bibinfo  {journal} {JHEP}\ }\textbf {\bibinfo {volume} {03}},\
  \bibinfo {pages} {127}},\ \Eprint {https://arxiv.org/abs/1709.03967}
  {arXiv:1709.03967 [hep-th]} \BibitemShut {NoStop}%
\bibitem [{\citenamefont {Hollands}(2018)}]{Hollands:2017chb}%
  \BibitemOpen
  \bibfield  {author} {\bibinfo {author} {\bibfnamefont {S.}~\bibnamefont
  {Hollands}},\ }\bibfield  {title} {\bibinfo {title} {Action principle for
  ope},\ }\href {https://doi.org/10.1016/j.nuclphysb.2017.11.013} {\bibfield
  {journal} {\bibinfo  {journal} {Nucl. Phys. B}\ }\textbf {\bibinfo {volume}
  {926}},\ \bibinfo {pages} {614} (\bibinfo {year} {2018})},\ \Eprint
  {https://arxiv.org/abs/1710.05601} {arXiv:1710.05601 [hep-th]} \BibitemShut
  {NoStop}%
\end{thebibliography}%

\ifarxiv

\appendix

\section{Conformal Blocks in the Null Square Limit}
\label{appSquareBlock}

In this appendix, we construct the null square limit of conformal blocks $\mathcal{F}(u_1,\dots,u_5)$ for the scalar five-point function $\langle \phi\phi\phi\phi\mathcal{O}\rangle$ that arises from two $\phi\times\phi$ OPEs~\cite{Bercini:2020msp}
\begin{align}
\label{eq:LCBlock}
    \mathcal{F}(u_i) &=
    \int_{0}^{1} dt_1dt_2\,\frac{\Gamma(2J_1+\tau_1)}{\Gamma\left(J_1+\frac{\tau_1}{2}\right)^2}\frac{\Gamma(2J_2+\tau_2)}{\Gamma\left(J_2+\frac{\tau_2}{2}\right)^2}
    \times \\ & \mspace{-30mu} \times
    (t_1(1-t_1))^{J_1+\frac{\tau_1-2}{2}}(t_2(1-t_2))^{J_2+\frac{\tau_2-2}{2}}
    \times \nonumber \\ & \mspace{-30mu} \times
    \frac{(1-t_1u_4 -u_2u_4+t_1u_2u_4)^{J_2-\ell}}{(1-t_1+t_1u_5)^{J_1-\ell+\frac{\Delta_\mathcal{O}+\tau_1-\tau_2}{2}}} u_1^{\frac{\tau_1}{2}}u_5^{\frac{\Delta_\mathcal{O}}{2}}
    \times \nonumber \\ & \mspace{-30mu} \times
    \frac{(1-t_2u_5 -u_2u_5+t_2u_2u_5)^{J_1-\ell}}{(1-t_2+t_2u_4)^{J_2-\ell+\frac{\Delta_\mathcal{O}+\tau_2-\tau_1}{2}}}u_3^{\frac{\tau_2}{2}}u_4^{\frac{\Delta_\mathcal{O}}{2}}
    \times \nonumber \\ & \mspace{-30mu} \times
    ((t_1+t_2-t_1t_2)(1-u_2)+u_2)^{-J_1-J_2+\frac{\Delta_{\mathcal{O}}-\tau_1-\tau_2}{2}}
    .
    \nonumber
\end{align}
This block is labeled by the twists $\tau_1,\tau_2$ and spins $J_1,J_2$ of the two exchanged fields $\mathcal{O}_1,\mathcal{O}_2$,  along with the integer $\ell=0,\dots,\mathrm{min}(J_1,J_2)$.  The latter denotes a basis of tensor structures $H_{12}^\ell V_{1,23}^{J_1-\ell} V_{2,31}^{J_2-\ell}$ for the spinning three-point function $\langle \mathcal{O}_1\mathcal{O}_2\mathcal{O}\rangle$. Throughout the appendix, we assume that the two exchanged fields have equal twist: $\tau_1=2h=\tau_2$.

The ordered null square limit $\mathrm{NS}_<$ is
\begin{equation}
x_{12}^2,x_{23}^2,x_{34}^2,x_{41}^2\rightarrow 0,\quad x_{12}^2,x_{34}^2 \ll x_{23}^2 \ll x_{41}^2.
\label{eq:app:ordered_NS_limit}
\end{equation}
For bookkeeping purposes, we define the above lightcone limits via infinitesimal rescalings $x_{ij}^2\rightarrow \epsilon_{ij} x_{ij}^2$, $\epsilon_{ij}\rightarrow 0$.  In this notation,  the five cross-ratios scale as
\begin{align}
&u_1 = O(\epsilon_{12}),\quad u_2 = O(\epsilon_{41}\epsilon_{23}),\quad u_3=O(\epsilon_{34}),\\
& u_4=O(\epsilon_{41}^{-1}),\quad u_5=O(\epsilon_{41}^{-1}).
\end{align}
In particular,  note that the ratio $x=u_4/u_5$ is finite in this limit. At leading order, the first two limits $x_{12}^2,x_{34}^2\rightarrow 0$ restrict the sum over descendants of $\mathcal{O}_1,\mathcal{O}_2$ to those with minimal twist $\Delta-J=2h$.  As a result, the leading asymptotics in these two limits is
\begin{equation}
\mathcal{F}(u_i) \sim (u_1u_3)^h (u_4u_5)^{h_\mathcal{O}} \tilde{\mathcal{F}}(u_2,u_4,u_5),
\end{equation}
where $\tilde{\mathcal{F}}(u_2,u_4,u_5)$ is a leading-twist block. To simplify future calculations, we  stripped off a $u_4u_5$-dependent prefactor and introduced the notation $2h_\mathcal{O}:= \Delta_\mathcal{O}$ for the half-twist of the external scalar $\mathcal{O}$. For the remaining two limits, the asymptotics of the leading-twist blocks $\tilde{\mathcal{F}}(u_2,u_4,u_5)$ are derived in two steps: First, we derive an integral representation for the most general solution to the Casimir equations. Next, we identify a basis of solutions consistent with this integral representation by analyzing a power series representation of the leading-twist blocks.

\paragraph{General solution to Casimir equations.} The Casi\-mir equations in cross-ratio space take the form
\begin{equation}
\tilde{\mathcal{D}}_a^2 \tilde{\mathcal{F}} = \left(J_a^2+O(J_i)\right) \tilde{\mathcal{F}},\quad a=1,2,
\end{equation}
where at leading order in the limit $\mathrm{NS}_<$, the differential operators $\tilde{\mathcal{D}}_a^2$ reduce to
\begin{align}
\mathcal{D}_a^2 =& \epsilon_{23}^{-1}\epsilon_{41}^{-1} \partial_2 \left(\vartheta_2-\vartheta_4-\vartheta_5-h_\mathcal{O} \right) \\&+ \epsilon_{23}^{-1} \partial_2 \partial_{6-a}+O(\epsilon_{23}^0). \nonumber
\end{align}
Here, we introduced the compact notation $\partial_i := \partial_{u_i}$ and $\vartheta_i := u_i \partial_{u_i}$ for $i=2,4,5$.  At leading order, we thereby obtain a simple system of two differential equations in three variables:
\begin{align}
&  \partial_2 \left(\vartheta_2-\vartheta_4-\vartheta_5-\frac{\Delta_\mathcal{O}}{2} \right) \tilde{\mathcal{F}}= \mathbf{J}^2 \tilde{\mathcal{F}} \\
& \partial_2 (\partial_4-\partial_5) \tilde{\mathcal{F}} = 2 \mathbf{j}^2 \tilde{\mathcal{F}}.
\end{align}
After applying the Laplace transform with respect to $u_2$,  it is easy to express the solutions to this system as integrals of a one-variable function:
\begin{align}
\tilde{\mathcal{F}}(u_2,u_4,u_5) =u_2^{h_\mathcal{O}} \int_0^\infty & \frac{d t}{t^{1+h_\mathcal{O}}}e^{-t-\frac{\mathbf{J^2} u_2}{t}-\mathbf{j}^2 u_2\frac{u_4-u_5}{t}}\nonumber  \\
& \times  \hat{f}\left(u_2 \frac{u_4+u_5}{t}\right).
\label{app:gen_sol_cas}
\end{align}
We have thereby reduced the problem to identifying a basis of functions $\hat{f}(Y)$ corresponding to the basis of tensor structures $\ell=0,\dots,\mathrm{min}(J_1,J_2)$.

\paragraph{Null square limit of leading-twist blocks.} Define  $(u_2,u_4,u_5):=(1-Z,v_2,v_1)$ and $\bar h_i:=h+J_i$.  For blocks in the $\ell$-basis of tensor structures, the following integral representation for blocks was derived in \cite{Kaviraj:2022wbw}:
\begin{align}
\tilde{\mathcal{F}}= Z^\ell \mspace{-10mu} \prod_{1\leq a\neq b\leq 2}& \frac{\Gamma(2\bar{h}_a)}{\Gamma(\bar{h}_a)^2} \int_0^1 \! {dt_a}  \frac{(t_a(1-t_a))^{\bar h_a-1} }{\left(1-(1-v_a)t_a\right)^{J_a-\ell+h_\mathcal{O}}} \nonumber \\
&  \times \frac{(1-(1-Z)v_a -Zv_at_b)^{J_a-\ell}}{(1-Z(1-t_1)(1-t_2))^{\bar h_{12;\mathcal{O}}}},
\end{align}
where $\bar h_{12;\mathcal{O}}:=\bar h_1+\bar h_2-h_\mathcal{O}$. The first limit $x_{23}^2\rightarrow 0$ corresponds to $Z\rightarrow 1^-$.  We analyze it by expanding the integrand around $Z=0$, resulting in a power series expansion of $\tilde{\mathcal{F}}$:
\begin{align}
\tilde{\mathcal{F}} &= \sum_{k=\ell}^\infty \frac{(\bar h_{12;\mathcal{O}})_k}{k!} \prod_{a=1}^2 \sum_{m_a=0}^{J_a-\ell} \binom{J_a-\ell-m_a}{m_a}
\times \\ & \mspace{140mu} \times
f_{k,m_1,m_2}(v_1,v_2) \, Z^{k+m_1+m_2}.\nonumber
\end{align}
The functions $f_{k,m_1,m_2}(v_1,v_2)$ can be determined explicitly in terms of a product of two Gauss hypergeometric functions with arguments $1-v_1,1-v_2$. Now,  for $Z=1+O(\epsilon_{23})$,  the derivative operator acts as $\partial_Z \tilde{\mathcal{F}}  = O(\epsilon_{23}^{-1}) \tilde{\mathcal{F}} $.  At the same time, the action of this derivative on each summand is $\brk{k+m_1+m_2}/{Z}$.  We deduce that the sum is dominated by the regime $k+m_1+m_2=O(\epsilon_{23}^{-1})$. Moreover, since $0\leq m_1,m_2\leq \mathrm{max}(J_1,J_2) = O(\epsilon_{23}^{-1/2})$, the powers $Z^{m_1+m_2}=1+O(\epsilon_{23}^{1/2})$ are trivial at leading order. This allows us to resum over $m_1,m_2$ and approximate the power series by
\begin{align}
\tilde{\mathcal{F}} \sim & \sum_{k=\ell}^\infty \frac{(\bar h_{12;\mathcal{O}})_k}{k!} Z^k\prod_{1\leq a\neq b\leq 2} \frac{(\bar h_a)_k}{(2\bar h_a)_k}\times \\& \times  F_1\left(\bar h_a; J_a-\ell-h_\mathcal{O},\ell-J_b;2\bar h_a+k;1-v_a,v_b \right),\nonumber
\end{align}
where $F_1$ denotes the Appell function of the first kind:
\begin{equation*}
F_1(b;a_1,a_2;c;z_1,z_2) = \prod_{a=1}^2 \sum_{n_a=0}^\infty \frac{(a_a)_{n_a}}{n_a!} z_a^{n_a} \frac{(b)_{n_1+n_2}}{(c)_{n_1+n_2}}.
\end{equation*}
Using this explicit expression for the power series, we can now approximate the region $k=O(\epsilon_{23}^{-1})$ that dominates the sum in the lightcone limit by an integral.  In this case, the integrand admits an expansion near $\epsilon_{23}=0$ with $J_1^2,J_2^2=O(\epsilon_{23}^{-1})$ and $0\leq \ell \leq \mathrm{min}(J_1,J_2)$.  At leading order,  the block therefore reduces to
\begin{multline}
\tilde{\mathcal{F}} \sim \mathcal{N}_{J_1J_2}^{h,h_\mathcal{O}} \int_0^\infty \frac{dk}{k^{1+h_\mathcal{O}}} e^{-k u_2-\frac{(J_1+J_2)^2}{2k}}
\times \\ \times
\mspace{-10mu}\prod\limits_{1\leq a\neq b\leq 2}\mspace{-10mu}
e^{- \frac{J_a}{k}\left(\frac{3J_a}{2} +(J_a-\ell)(1-v_a)-(J_b-\ell)v_b\right)}
\,,
\end{multline}
where
\begin{equation}
\mathcal{N}_{J_1J_2}^{h,h_\mathcal{O}}:=\frac{1}{\Gamma(J_1+J_2+2h-h_\mathcal{O})} \prod_{a=1}^2 \frac{\Gamma(2J_a+2h)}{\Gamma(J_a+h)}.
\label{norm_blocks_NS}
\end{equation}
The following approximation of blocks was based solely on the lightcone limit $x_{23}^2\rightarrow 0$, where the cross-ratios $u_4,u_5=v_2,v_1$ remained finite. In the final lightcone limit $x_{41}^2\rightarrow 0$, the latter scale as $v_a = O(\epsilon_{41}^{-1})$.  We would like to identify the blocks in this limit with the integral representation~\eqref{app:gen_sol_cas} of solutions to the Casimir equations by changing variables to $t:= k u_2$. To obtain a basis of blocks that is consistent with the functional form of eq.~\eqref{app:gen_sol_cas},  we assume that the $\ell$-dependent terms in the exponential remain as leading contributions,
\begin{equation}
J_av_b u_2(J_b-\ell)  = O(1) \iff J_b-\ell = O\brk!{\epsilon_{23}^{-{1}/{2}} \epsilon_{41}^{{1}/{2}}}.
\end{equation}
Changing variables according to~\arxor{\eqref{eqCasimirVar}}{(8)} from $(J_1,J_2,\ell)$ to $(J,j_1,j_2)$, allows us to parameterize the large-spin limit as
\begin{equation}
J^2 = O(\epsilon_{23}^{-1}\epsilon_{41}^{-1})\,, \quad \, j_a^2 = O(\epsilon_{23}^{-1}\epsilon_{41}).\,
\end{equation}
After expanding the Gamma functions of $\mathcal{N}_{J_1J_2}^{h,h_\mathcal{O}}$ in eq.~\eqref{norm_blocks_NS}, we finally obtain
\begin{multline}
       \tilde{\mathcal{F}}=
         2^{2h+2J+h_\mathcal{O}-1}\pi^{-1/2}J^{{1}/{2}+h_\mathcal{O}}
       u_2^{h_\mathcal{O}}
       \times \\ \times
       \int_0^\infty
       \frac{dt
       }{
         t^{1+h_\mathcal{O}}
       }
       e^{-t-\frac{J^2 u_2}{t}-\frac{J j_1u_2u_4}{t}-\frac{J j_2u_2u_5}{t}}
       \,.
       \label{eq:NullSquareBlockResult}
\end{multline}
This expression coincides with~\arxor{\eqref{eqNullBlockMain}}{(7)} after integrating over~$t$, in addition to setting $2h=2+\gamma$ for the twist of the exchanged fields and $2h_\mathcal{O}=\Delta_\mathcal{L}$ for the twist (scaling dimension) of the fifth external scalar.

\section{Null Square Inversion Formula}
\label{appInversion}

This appendix is divided into two parts: First, we invert the conformal block decomposition of the five-point function in the ordered null-square limit $NS_<$ of eq.~\eqref{eq:app:ordered_NS_limit}. Next, by specializing this inversion formula to five-point functions that factorize in the null square limit, we demonstrate that $C_\mathcal{O}(J_1,J_2,\ell)$ reduces to a one-variable function of the ratio $r=(J_2-\ell)/(J_1-\ell)$, thereby proving the uniqueness of eq.~\arxor{\eqref{eqGFactorized}}{(15)}.
\paragraph{Derivation of the inversion formula.} The derivation is based on the observation that the null square blocks in eq.~\eqref{app:gen_sol_cas} are the integral transform of a simple power-law-times-exponential function.  After the change of variables $t:=ku_2$, we can identify this integral as a straightforward generalization of the Laplace transform, which we denote by
\begin{equation}
G(u_2,u_4,u_5) =: \int_0^\infty dk\, e^{-k u_2} \mathbf{L}^{-1}[G]\left(k,\frac{u_4}{k},\frac{u_5}{k}\right).
\label{eq:app:Laplace5}
\end{equation}
In this case, we can express the inverse Laplace transform of the five-point blocks in the ordered null-square limit as
\begin{equation}
\mathbf{L}^{-1}[\tilde{\mathcal{F}}](k,w_4,w_5) = \frac{\mathcal{N}_5(J)}{k^{1+h_\mathcal{O}}} e^{-J\left(\frac{J}{k}+j_1w_4+j_2w_5\right)},
\end{equation}
where $w_4=u_4/k$, $w_5=u_5/k$, $h_\mathcal{O}=\Delta_\mathcal{O}/2$, and
\begin{equation}
\mathcal{N}_5(J):=2^{2h+2J+\Delta_\mathcal{O}/2-1} J^{1/2+\Delta_\mathcal{O}/2} \pi^{-1/2}.
\end{equation}
Given that the five-point function reduces to leading-twist exchange $G_\mathcal{O} \sim (u_1u_3)^h (u_4u_5)^{h_\mathcal{O}}\tilde{G}_\mathcal{O}(u_2,u_4,u_5)$ in the limit $u_1,u_3\rightarrow 0$, we can then express the null square conformal block decomposition as
\begin{align}
&\mathbf{L}^{-1}[\tilde{G}_{\mathcal{O}}](k,w_4,w_5) = \nonumber\\
& \int_0^\infty \mspace{-10mu} d(J^2) e^{-J^2 k^{-1}} \int_0^\infty \mspace{-10mu} d(J j_1) e^{-Jj_1 w_4} \int_0^\infty \mspace{-10mu} d(J j_2) e^{-J j_2 w_5} \nonumber\\
& \times\frac{\mathcal{N}_5(J)}{8 J^3 k^{1+h_\mathcal{O}}} C(J)^2 C_\mathcal{O}(J,j_1,j_2),
\end{align}
where the original measure is $dJ_1 dJ_2 d\ell/4$, with a factor of four to account for even spin exchange in the two OPEs. Now, in this $k$-space, the conformal block decomposition itself reduces to another series of Laplace transforms with respect to $(J^2, Jj_1,Jj_2)$.  After applying their inverse transforms, we obtain
\begin{multline}
C_\mathcal{O}(J,j_1,j_2)=
\frac{8 J^3}{\mathcal{N}_5(J)C(J)^2}\int_{(c+i\mathbb{R})^3} \frac{dk}{k^{1-h_\mathcal{O}}} dw_4 dw_5
\\ \times
e^{J^2\!/k + J j_1 w_4+ J j_2 w_5}\,\mathbf{L}^{-1}[\tilde{G}_{\mathcal{O}}](k,w_4,w_5)
\,.
\label{eq:app:lif5}
\end{multline}
It is now straightforward to write down an inversion formula for the position space correlator by inserting the formula for the inverse Laplace transform with respect to $k$:
\begin{equation}
\mathbf{L}^{-1}[\tilde{G}_{\mathcal{O}}](k,w_4,w_5) = \int_{c+i\mathbb{R}} d u_2 e^{k u_2} \tilde{G}_{\mathcal{O}}(u_2,kw_4,kw_5).
\label{eq:app:invLaplace5}
\end{equation}
Here, following the standard definition of the inverse Laplace transform, $c>0$ is a constant shift of the contours of integration to the right of all poles and branch cuts of the integrand in the complex plane. As a result, we obtain the following inversion formula for the five-point function in the ordered null square limit:
\begingroup
\setlength{\widefboxpadding}{.2em}
\begin{empheq}[box=\widefbox]{align}
& C_\mathcal{O}(J,j_1,j_2)=
\frac{2^{4-2h-\Delta_\mathcal{O}/2} J^{\frac{5-\Delta_\mathcal{O}}{2}}}{\hat{C}(J)^2}
\nonumber \\ & \times
\mspace{-10mu} \int\limits_{(c+i\mathbb{R})^4} \mspace{-10mu} \frac{dk\,du_2\,dw_4\,dw_5}{k^{1+\Delta_\mathcal{O}/2}}
\,e^{ku_2 + J^2\!/k + J j_1 w_4+ J j_2 w_5}
\nonumber \\ & \mspace{50mu} \times
\frac{
G_{\mathcal{O}}(u_1,u_2,u_3,u_4=kw_4,u_5=kw_5)
}{(u_1u_3)^{h}(w_4w_5)^{\Delta_\mathcal{O}/2}}
\,.
\label{eq:InversionFormula}
\end{empheq}
\endgroup

\paragraph{Consequences of factorization for OPE coefficients.} We now consider five-point functions with the factorization property
\begin{equation}
G_\mathcal{O}(u_1,\dots,u_5) \sim G_4(u_1u_3,u_2) (u_4u_5)^{h_\mathcal{O}} f_\mathcal{O}(u_4,u_5),
\label{eq:app:factorized5pt}
\end{equation}
where $G_4(u,v)$ is the four-point function in the null square limit, while $f_\mathcal{O}$ is a symmetric function of two variables that is homogeneous of degree $-h_{\mathcal{O}}$:
\begin{equation}
f_\mathcal{O}(\lambda u_5,\lambda u_4)=f_\mathcal{O}(\lambda u_4,\lambda u_5) = \lambda^{-h_\mathcal{O}}f_\mathcal{O}(u_4,u_5).
\end{equation}
Given $G_4(u,v) \sim u^h g_4(v)$ in the ordered null square limit, the inverse Laplace transform~\eqref{eq:app:invLaplace5} of a factorized function~\eqref{eq:app:factorized5pt} will itself factorize as well:
\begin{equation*}
\mathbf{L}^{-1}[g_4\, f_\mathcal{O}](k,w_4,w_5) = k^{-h_\mathcal{O}} \mathbf{L}^{-1}[g_4](k)\, f_\mathcal{O}(w_4,w_5).
\end{equation*}
We can therefore separate the integral over $k$ from the integrals over $w_4,w_5$ in the inversion formula~\eqref{eq:app:invLaplace5}. In doing so, we identify the inversion formula for the four-point OPE coefficients $C(J)^2$:
\begin{equation}
C(J)^2 = \frac{4J}{\mathcal{N}_4(J)} \int_{c+i\mathbb{R}} \frac{dk}{k} e^{\frac{J^2}{k}} \mathbf{L}^{-1}[g_4](k),
\end{equation}
where $\mathcal{N}_4(J)=4^{J+h}J^{1/2}\pi^{-1/2}$, which follows from the Laplace transform with respect to $v$ of the Bessel function $K_0(2J\sqrt{v})$ in four-point conformal blocks. Given $\mathcal{N}_5(J)=2^{h_\mathcal{O}-1}J^{h_\mathcal{O}}\mathcal{N}_4(J)$, the inversion formula therefore reduces to
\begin{multline}
\left(\frac{J}{2}\right)^{h_\mathcal{O}-2}C_\mathcal{O}(J,j_1,j_2) =\\
\int_{(c+i\mathbb{R})^2} dw_4 dw_5 e^{Jj_1w_4+Jj_2 w_5} f_\mathcal{O}(w_4,w_5).
\end{multline}
The RHS of this equation is the inverse Laplace transform of $f_\mathcal{O}$ with respect to each of its arguments $w_4,w_5$. Since the latter function is homogeneous of degree $-h_\mathcal{O}$, then the LHS (\ie the Laplace transform of $f_\mathcal{O}$) must be a homogeneous function of degree $h_\mathcal{O}-2$ in $(Jj_1,Jj_2)$. As a result, factorization in the null square limit implies that OPE coefficients take the most general form
\begin{equation}
C_\mathcal{O}(J,j_1,j_2) = j_1^{h_\mathcal{O}-2} C_\mathcal{O}(r),\,\,\, r=j_2/j_1,
\end{equation}
in agreement with eq.~\arxor{\eqref{eqAssumeChat}}{(14)} for the OPE coefficients $C_{\mathcal{L}}$ normalized by their tree-level value $C_{\mathcal{L}}^{(0)} = 8$. Finally, after re-inverting the relation between $C_\mathcal{O}$ and $f_\mathcal{O}$ to
\begin{equation}
f_\mathcal{O}(u_4,u_5) = \int_{\mathbb{R}_+^2} dj_1 dj_2 e^{-(j_1 u_4+j_2 u_5)} \left(\frac{j_1}{2}\right)^{h_\mathcal{O}-2}C_\mathcal{O}(r),
\end{equation}
we can explicitly integrate over $r$ by parameterizing the homogeneous function as
\begin{equation}
f_{\mathcal{O}}(u_4,u_5) = (u_4 u_5)^{-h_\mathcal{O}/2} F_\mathcal{O}(x),\,\,\, x=u_4/u_5.
\end{equation}
As a result, the null square conformal block decomposition reduces to
\begin{equation}
F_\mathcal{O}(x) = \frac{\Gamma(h_\mathcal{O}-1)}{2^{h_\mathcal{O}-2}}x^{h_\mathcal{O}/2} \int_0^\infty \frac{dx}{(x+r)^{h_\mathcal{O}}} C_\mathcal{O}(r),
\end{equation}
in agreement with eq.~\arxor{\eqref{eqFDerivation}}{(16)} for $\mathcal{O}=\mathcal{L}$.

\section{Three-Loop Results}
\label{app3Loops}

The weak-coupling expressions for the finite function $\hat{F}(x)$ and
the structure constant $\hat{C}_\mathcal{L}$ are given in terms of
harmonic polylogarithms (HPLs). These functions are defined
recursively, via
\begin{equation}
    \label{Hlogrecursion}
    H_{a_1,a_2,\dots,a_n}(x) = \int_{0}^{x}\frac{dz}{z-a_1}H_{a_2,\dots,a_n}(z)\,
\end{equation}
with the seed $H(x)=1$ and $a_i \in \{-1,0,1\}$. We use the compact HPL notation introduced in \cite{Remiddi:1999ew}, in which a string of $n-1$ zero indices followed by $\pm 1$ is replaced by $\pm n$, \ie $H_{3,0} = H_{0,0,1,0}(x)$.

The three-loop contribution to $\hat{F}(x)$ is given by
\begin{align}
    \hat{F}^{(3)}(x)=&16\zeta_3H_{-2-1}+32\zeta_3H_{-20}+16\zeta_3H_{-1-2}-\nonumber\\
    -&32\zeta_3H_{-1-1-1}+16\zeta_3H_{-1-10}+16\zeta_3H_{-100}-\nonumber\\
    -&8\zeta_3H_{000}-144\zeta_2H_{-3-1}+88\zeta_2H_{-30}-\nonumber\\
    -&96\zeta_2H_{-2-2}-96\zeta_2H_{-1-3}-516\zeta_4H_{-1-1}+\nonumber\\
    +&360\zeta_4H_{-10}-646\zeta_4H_{00}+48\zeta_2H_{-2-1-1}-\nonumber\\
    -&32\zeta_2H_{-2-10}+88\zeta_2H_{-200}+48\zeta_2H_{-1-2-1}-\nonumber\\
    -&48\zeta_2H_{-1-20}+48\zeta_2H_{-1-1-2}-\nonumber\\
    -&96\zeta_2H_{-1-1-1-1}+48\zeta_2H_{-1-1-10}-\nonumber\\
    -&96\zeta_2H_{-1-100}+96\zeta_2H_{-1000}-216\zeta_2H_{0000}+\nonumber\\
    +&48H_{-400}-48H_{-3-100}+40H_{-3000}-\nonumber\\
    -&32H_{-2-200}-32H_{-1-300}+16H_{-2-1-100}-\nonumber\\
    -&16H_{-2-1000}+40H_{-20000}+16H_{-1-2-100}-\nonumber\\
    -&32H_{-1-2000}+16H_{-1-1-200}-\nonumber\\
    -&32H_{-1-1-1-100}+16H_{-1-1-1000}-\nonumber\\
    -&48H_{-1-10000}+48H_{-100000}-120H_{000000}-\nonumber\\
    -&40\zeta_3^2-48\zeta_3H_{-3}-8\zeta_2\zeta_3H_0+144\zeta_2H_{-4}+\nonumber\\
    +&394\zeta_4H_{-2}-112\zeta_5H_{-1}+32\zeta_5H_0-\nonumber\\
    -&\frac{3085}{2}\zeta_6+\frac{1}{N^2}\Big(-96\zeta_3H_{-20}-96\zeta_3H_{-100}+\nonumber\\
    +&192\zeta_3H_{000}+48\zeta_2H_{-30}+144\zeta_2H_{-2-2}+\nonumber\\
    +&144\zeta_2H_{-1-3}-1620\zeta_4H_{-1-1}+312\zeta_4H_{-10}+\nonumber\\
    +&300\zeta_4H_{00}-48\zeta_2H_{-1-20}-432\zeta_2H_{-1-100}+\nonumber\\
    +&96\zeta_2H_{-1000}+48\zeta_2H_{0000}+48H_{-3000}+\nonumber\\
    +&48H_{-2-200}+48H_{-1-300}+96H_{-20000}-\nonumber\\
    -&48H_{-1-2000}-432H_{-1-10000}+\nonumber\\
    +&96H_{-100000}+144H_{000000}-528\zeta_3^2-231\zeta_6+\nonumber\\
    +&384\zeta_2\zeta_3H_{-1}+192\zeta_2\zeta_3H_0-216\zeta_4H_{-2}+\nonumber\\
    +&48\zeta_5H_{-1}+288\zeta_5H_0+\nonumber\\
    +&\frac{1}{1+x}\Big(-48\zeta_2H_{-30}+288\zeta_2H_{-200}+\nonumber\\
    +&144\zeta_2H_{-1000}-48\zeta_2H_{0000}+180\zeta_4H_{-10}-\nonumber\\
    -&300\zeta_4H_{00}-192\zeta_3H_{000}-96H_{-400}-\nonumber\\
    -&48H_{-3000}+192H_{-20000}+240H_{-100000}-\nonumber\\
    -&144H_{000000}-2217\zeta_6-288\zeta_2H_{-4}-\nonumber\\
    -&480\zeta_3\zeta_2H_0+1440\zeta_4H_{-2}-432\zeta_5H_0\Big)\Big).
\end{align}

While the three-loop contribution to the structure constant
$\hat{C}_\mathcal{L}$ is a much simpler function, given by
\begin{align}
    \hat{C}_{\mathcal{L}}^{(3)}(r)  &=16\zeta_3H_{20}-16\zeta_3H_{21}+8\zeta_3H_{000}+196\zeta_4H_{00}+\nonumber\\
    &+16\zeta_2H_{22}+48\zeta_2H_{30}+16\zeta_2H_{31}+48\zeta_2H_{200}+\nonumber\\
    &+48\zeta_2H_{210}+32\zeta_2H_{211}+96\zeta_2H_{0000}+48H_{50}+\nonumber\\
    &+32H_{230}+32H_{320}+40H_{400}+48H_{410}+\nonumber\\
    &+16H_{2120}+32H_{2200}+16H_{2210}+40H_{3000}+\nonumber\\
    &+16H_{3100}+16H_{3110}+48H_{20000}+48H_{21000}+\nonumber\\
    &+16H_{21100}+32H_{21110}+120H_{000000}-24\zeta_3^2-\nonumber\\
    &-1079\zeta_6+32\zeta_3H_3+48\zeta_2H_4+156\zeta_4H_2-\nonumber\\
    &-32\zeta_5H_0+\frac{1}{N^2}\Big(-96\zeta_3H_{20}+192\zeta_3H_{100}+\nonumber\\
    &+288\zeta_4H_{10}+96\zeta_2H_{120}-96\zeta_2H_{200}-\nonumber\\
    &-96\zeta_2H_{1000}-96\zeta_2H_{1100}-96H_{50}-96H_{140}-\nonumber\\
    &-48H_{230}-48H_{320}-48H_{1300}+48H_{2200}+\nonumber\\
    &+288H_{3000}+192H_{12000}+336H_{20000}+\nonumber\\
    &+432H_{21000}+144H_{100000}+240H_{110000}-\nonumber\\
    &-624\zeta_3^2-528\zeta_6-96\zeta_3H_3+288\zeta_2\zeta_3H_0+\nonumber\\
    &+288\zeta_2\zeta_3H_1+144\zeta_4H_2+144\zeta_5H_0+432\zeta_5H_1\Big)\,.\label{eqPertubativeB3}
\end{align}

Note that the $\hat{F}^{(3)}(x)$ contains terms with a rational prefactor multiplying the HPLs, while the structure constant is simply a linear combination of HPLs. This happens because in the
inversion formula~\arxor{\eqref{eqInversion}}{(17)}, one divides the discontinuity
by $x$ and then integrates. Once we integrate the terms with $x$ or $1+x$ in the denominator times a HPL, via the very definition of these
functions \eqref{Hlogrecursion} we get another HPL but with different weight.

\section{Null Pentagon Limit}
\label{appNullPent}

The null pentagon limit can be achieved by first taking
$x_{12}^2,x_{34}^2 \to 0$ (or $u_1,u_3\to0$), projecting to
leading-twist operators in the OPE. Further taking
$x_{45}^2,x_{15}^2\to0$ (or $u_4,u_5\to0$),
large-spin operators dominate. At this stage, the
polarization $\ell$ is still finite, but by taking the last distance to become null, $x_{23}^2\to0$ (or $u_2\to0$), we project also to large $\ell$. The conformal block in the pentagon limit ($u_i \to 0$) simplifies dramatically, and is given by a simple
exponential~\cite{Bercini:2020msp}:
\begin{multline}
    \mathcal{F}(u_i)=
    2^{3+\gamma_1+\gamma_2}J_1^{1-\frac{\gamma_2}{2}}J_2^{1-\frac{\gamma_1}{2}}
    \ell^{-2+\gamma_1+\gamma_2}
    \times\\ \times
    u_1^{\frac{2+\gamma_1}{2}}u_3^{\frac{2+\gamma_2}{2}}u_4^2u_5^2
    \,e^{-\ell u_2 - \frac{J_2^2u_4}{\ell}- \frac{J_1^2u_5}{\ell}}\,,
    \label{eqBlockNullPentagon}
\end{multline}
where $\gamma_i=\gamma(J_i)$ are the anomalous dimensions of the two
exchanged operators.

Notice that this limit is very different than the null square limit
considered in the main text. Once we take the five neighboring
distances to become null separated, the null pentagon correlation
function has no finite cross-ratios. In terms of the quantum numbers,
this limit is approached by first taking the spins $J_i$ to be large,
and then the polarization $\ell$, hence there are also no finite
ratios of the quantum numbers in the null pentagon limit.

The conformal block in this limit is independent of the fifth
external operator, so any difference in the
correlation functions must come from the different three-point
functions of the block decomposition~\arxor{\eqref{eqStartRelation}}{(3)}.
Conversely, the equality between the correlators~\arxor{\eqref{eqDuality}}{(24)}
implies that their tree-level normalized structure
constants must also be identical:
\begin{equation}
    \hat{C}_{\phi}(J_1,J_2,\ell) = \hat{C}_{\mathcal{L}}(J_1,J_2,\ell)\,.
    \label{eqDualityC}
\end{equation}

The null pentagon correlator $\hat{G}_\phi$ (or $\hat{G}_\mathcal{L}$)
must be cyclically symmetric (\ie invariant under $u_i \to u_{i+1}$). By
demanding this symmetry of the correlator, one can bootstrap the
universal behavior of the structure constants.
This was done for
$\hat{C}_\phi$ in~\cite{Bercini:2020msp}, which, due
to~\eqref{eqDualityC}, immediately gives the following result:
The three-point function of two
leading-twist large-spin operators and the Lagrangian in the limit of
large $\ell$ are
\begin{multline}
    \hat{C}_{\mathcal{L}}(J_1,J_2,\ell) =
    \\
    \mathcal{N}(\lambda)\,
    e^{-\frac{f(\lambda)}{4}(\log{\ell}^2+2\log{2}\log{(J_1J_2)})-\frac{g(\lambda)}{2}\log{\ell}}
    \,,
    \label{eqCNullPentApp}
\end{multline}
where $\mathcal{N}(\lambda)$ is a coupling-dependent but
spin-in\-de\-pen\-dent factor that bootstrap arguments cannot fix.

\section{Trivial Relation}
\label{appTrivial}

In the following, we want to show that
\begin{align}
\mathcal{I}\left[ x\int_{0}^{\infty} dr \, \frac{\hat{C}_{\mathcal{L}}(r)}{(x+r)^2}\right] = \hat{C}_{\mathcal{L}}(1)\label{eqAppTrivial}
\end{align}
is trivially satisfied for any physical structure constant.

The first step is to use the fact that $\hat{C}_{\mathcal{L}}(r)$ is
invariant under the inversion $r \to 1/r$ to write the single
integral above as the sum of two integrals
\begin{equation}
    \int_{0}^{\infty} dr \, \frac{\hat{C}_{\mathcal{L}}(r)}{(x+r)^2} =
    \int_{0}^{1}dr\left(\frac{\hat{C}_{\mathcal{L}}(r)}{(x+r)^2}+\frac{\hat{C}_{\mathcal{L}}(r)}{(1+xr)^2}\right)
    \,.
\end{equation}
The advantage of this step is that now it is clear that the integral
of any polynomial in $r$ is convergent.

To complete our derivation, we note that the structure constant
$\hat{C}_{\mathcal{L}}(r)$ better be regular around $r=1$, since at
this value it is equal to the cusp anomalous
dimension~\arxor{\eqref{eqCusptoB}}{(30)}. Therefore, we
can Taylor expand the structure constant around this point
\begin{equation}
    \hat{C}_{\mathcal{L}}(r) = \sum_{n=0}^\infty c_n (r-1)^n
\end{equation}
and plug into the initial relation~\eqref{eqAppTrivial} to obtain
\begin{equation}
    \sum_{n=0}^\infty c_n\mathcal{I}\left[x\int_{0}^{1}dr\brk*{\frac{(r-1)^n}{(x+r)^2}+\frac{(r-1)^n}{(1+xr)^2}} \right] = c_0\,.
\end{equation}
Performing the integral and applying the functional for the first term
of the sum $n=0$ allows us to simplify the relation above into the
sum rule
\begin{equation}
    \sum_{n=1}^\infty c_n\mathcal{I}\left[x\int_{0}^{1}dr\left(\frac{(r-1)^n}{(x+r)^2}+\frac{(r-1)^n}{(1+xr)^2}\right) \right] = 0\,.
    \label{eqAppSumRule}
\end{equation}

Therefore, the relation~\eqref{eqAppTrivial} will be trivially
satisfied if each term of the sum~\eqref{eqAppSumRule} is identically
zero. It turns out that the integrals and the functional are simple
enough to check this explicitly:
\begin{align}
    \mathcal{I}&\left[x\int_{0}^{1}dr\left(\frac{(r-1)^n}{(x+r)^2}+\frac{(r-1)^n}{(1+xr)^2}\right)
    \right]
    \nonumber \\ & \mspace{20mu}
    = (-1)^n\left(1+\sum_{k=0}^n k
    \binom{n}{k}\mathcal{I}[x^k\log{x}]\right)
    \nonumber \\ & \mspace{20mu}
    = (-1)^n\left(1+\sum_{k=0}^n k \binom{n}{k}\frac{(-1)^k}{k}\right) =0\,,
\end{align}
where in the last line we used the fact that
\begin{align}
    \mathcal{I}[x^p\log^q(x)] &=\lim_{\epsilon \to 0}\frac{\partial^q}{\partial\epsilon^q} \mathcal{I}[x^{p+\epsilon}] = \\
    &=\frac{q!}{p^q}\sum_{k=0}^{q}\frac{(-1)^{q+p-1}(\pi p)^{k-1}}{k!}\sin\left(\frac{\pi k}{2}\right)\nonumber\,.
\end{align}

\fi

\end{document}